\documentclass[journal]{IEEEtran}

\usepackage{amsmath}
\usepackage{amssymb}
\usepackage{amsfonts}
\usepackage{graphicx}
\usepackage{epstopdf}
\usepackage{subfigure} 
\usepackage{color}
\usepackage{bm}
\usepackage{url}

\usepackage{cite}
\usepackage{balance}
\usepackage{microtype}

\usepackage{pgfplots}
\pgfplotsset{compat=newest}
\usetikzlibrary{plotmarks}
\usetikzlibrary{arrows.meta}
\usepgfplotslibrary{patchplots}
\usepackage{grffile}

\ifCLASSINFOpdf
\else
   \usepackage[dvips]{graphicx}
\fi
\usepackage[ruled,boxed,linesnumbered,commentsnumbered]{algorithm2e}
\usepackage{xcolor}

\begin{document}

\title{Deep-Unfolded Massive Grant-Free Transmission \\ in Cell-Free Wireless Communication Systems}

\author{Gangle~Sun,~\IEEEmembership{Student~Member,~IEEE}, Mengyao~Cao, Wenjin~Wang,~\IEEEmembership{Member,~IEEE}, \\ Wei~Xu,~\IEEEmembership{Fellow,~IEEE}, and Christoph~Studer,~\IEEEmembership{Senior~Member,~IEEE}
\thanks{This work was supported in part by the National Natural Science Foundation of China under Grant 62371122 and Grant 62341110, in part by the National Key Research and Development Program under Grant 2020YFB1806608, and in part by the Special Fund for Key Basic Research in Jiangsu Province No.~BK20243015. 
The work of C. Studer was supported in part by an ETH Research Grant.}
\thanks{Part of this work was presented at the IEEE International Workshop on Signal Processing Advances in Wireless Communications (SPAWC) 2023~\cite{sun2023joint_arxiv} and the 19th International Symposium on Wireless Communication Systems (ISWCS)~\cite{sun2023deep_unfolding}, respectively. 
This work extends our previous results by (i) unifying the JACD optimization problem in~\cite{sun2023joint_arxiv} and the JAD optimization problem in~\cite{sun2023deep_unfolding}; (ii) proposing a deep-unfolding-based approximate box-constrained algorithm, in which complex computations based on the KKT conditions in~\cite{sun2023joint_arxiv} are replaced by a simple shrinkage operation; (iii) implementing a novel model training strategy that facilitates the joint training of parameters in the FBS modules and the AUD module, fully utilizing the estimated channel information; and (iv) providing a comprehensive performance assessment of different algorithms across various scenarios.
The associate editor coordinating the review of this article and approving it for publication was Prof. Andreas Burg. \textit{(Corresponding authors: Wenjin Wang; Christoph Studer.)}}
\thanks{Gangle~Sun, Mengyao~Cao, Wenjin~Wang, and Wei~Xu are with the National Mobile Communications Research Laboratory, Southeast University, Nanjing 210096, China, and also with the Purple Mountain Laboratories,
Nanjing 211100, China (e-mail: sungangle@seu.edu.cn; cmengyao@seu.edu.cn; wangwj@seu.edu.cn; wxu@seu.edu.cn). }
\thanks{Christoph Studer is with the Department of Information Technology and Electrical Engineering, ETH Zurich, 8092 Z\"urich, Switzerland (e-mail:
studer@ethz.ch).}
}

\maketitle

\begin{abstract}
Grant-free transmission and cell-free communication are vital in improving coverage and quality-of-service for massive machine-type communication. 
This paper proposes a novel framework of joint active user detection, channel estimation, and data detection (JACD) for massive grant-free transmission in cell-free wireless communication systems. 
We formulate JACD as an optimization problem and solve it approximately using forward-backward splitting. 
To deal with the discrete symbol constraint, we relax the discrete constellation to its convex hull and propose two approaches that promote solutions from the constellation set. 
To reduce complexity, we replace costly computations with approximate shrinkage operations and approximate posterior mean estimator computations. 
To improve active user detection (AUD) performance, we introduce a soft-output AUD module that considers both the data estimates and channel conditions. 
To jointly optimize all algorithm hyper-parameters and to improve JACD performance, we further deploy deep unfolding together with a momentum strategy, resulting in two algorithms called DU-ABC and DU-POEM.
Finally, we demonstrate the efficacy of the proposed JACD algorithms via extensive system simulations. 
\end{abstract}

\begin{IEEEkeywords}
Active user detection, cell-free communication, channel estimation, data detection, deep unfolding, grant-free transmission, massive machine-type communication.
\end{IEEEkeywords}

\IEEEpeerreviewmaketitle

\section{Introduction}\label{Section_I}

\IEEEPARstart{M}{assive} machine-type communication (mMTC) is an essential scenario of fifth-generation (5G) and beyond-5G wireless communication systems~\cite{ITU2023framework, Sun2024beam, gao2022grant, Mei2022compressed, Sun2024hybrid, gao2022joint}, and 
focuses on supporting a large number of sporadically active user equipments (UEs) transmitting short packets to an infrastructure base station (BS)~\cite{andrews2014what, Fang2024OFDMA,chen2021massive, Sun2022Hybrid, sun2024low, Bjornson2017arandom,xu2023toward, Sun2022massive}. 
On the one hand, grant-free transmission schemes~\cite{Liu2018sparse, zhang2016grant, sun2021OFDMA, zhu2022massive} are essential for mMTC scenarios, as they reduce signaling overhead, network congestion and transmission latency compared with traditional grant-based transmission schemes~\cite{Liu2018massive,chen2018sparse,zhu2022OFDM}. 
In grant-free transmission schemes, UEs transmit signals directly to the BS over shared resources, bypassing the need for complex scheduling.
On the other hand, to improve coverage in mMTC, cell-free communication offers a promising solution~\cite{bjornson2020scalable,ngo2017cell, Mishra2022rate, Ngo2015cell, Gao2024compressed}. 
In cell-free systems, numerous decentralized access points (APs) are connected to a central processing unit (CPU), jointly serving UEs to effectively broaden the coverage, mitigate inter-cell interference and enhance spectral efficiency~\cite{elhoushy2021cell,zhang2019cell}.
The key tasks of the CPU for massive grant-free transmission in cell-free wireless communication systems involve (i) identifying the set of active UEs, (ii) estimating their channels, and (iii) detecting their transmitted data.

\subsection{Contributions}
This paper proposes a novel framework for joint active user detection, channel estimation, and data detection (JACD) in cell-free systems with grant-free access.
We start by formulating JACD as an optimization problem that fully exploits the sparsity of the wireless channel and data matrices and then approximately solve it using forward-backward splitting (FBS)~\cite{goldstein2014field,beck2009fast}.
To enable FBS, we relax the discrete constellation constraint to its convex hull and employ JACD methods with the incorporation of either a regularizer or a posterior mean estimator (PME), guiding symbols toward discrete constellation points, resulting in the box-constrained FBS and PME-based JACD algorithms, respectively. 
To reduce complexity, we replace the exact proximal operators with approximate shrinkage operations and approximate PME computations. 
To improve convergence and JACD performance, we include per-iteration step sizes and a momentum strategy.
To avoid tedious manual parameter tuning, we employ deep unfolding (DU) to jointly tune all of the algorithm hyper-parameters using machine learning tools.
To improve active user detection (AUD) performance, we include a novel soft-output AUD module that jointly considers the estimated data and channel matrix.
Based on the aforementioned modifications, we have developed the deep unfolding versions of the box-constrained FBS and PME-based JACD algorithms, referred to as DU-ABC and DU-POEM.
We use Monte--Carlo simulations to demonstrate the superiority of our framework compared to existing methods.

\subsection{Prior Work}
\subsubsection{Massive Grant-Free Transmission in Cell-Free Wireless Communication Systems}
Recent results have focused on AUD, channel estimation (CE), and data detection (DD) for massive grant-free transmission in cell-free wireless communication systems \cite{sun2023joint_arxiv,sun2023deep_unfolding, Ganesan2021algorithm, shao2020covariance, wang2021grant, Li2023Asynchronous, Diao2023Scalable, Jiang2023EM,wang2022two,guo2022joint, Ke2021massive, Johnston2022model, Di2022adaptive, Femenias2023repetition, zhou2023active, Iimori2021grant}.
Reference~\cite{Ganesan2021algorithm} proposes two different AUD algorithms based on dominant APs and clustering, respectively. On this basis, a parallel AUD algorithm is developed to reduce complexity.
Reference~\cite{shao2020covariance} proposes a covariance-based cooperative AUD method in which APs exchange their low-dimensional local information with neighbors.
Reference~\cite{wang2021grant} introduces a correlation-based AUD algorithm, accompanied by simulation results and empirical analysis, demonstrating that the cell-free system outperforms the collocated system in AUD performance.
Reference~\cite{Li2023Asynchronous} proposes centralized and distributed AUD algorithms for asynchronous transmission caused by low-cost oscillators.
For near-real-time transmission, reference~\cite{Diao2023Scalable} introduces a deep-learning-based AUD algorithm, in which distributed computing units employ convolutional neural networks for preliminary AUD, and the CPU subsequently refines them through transfer learning.
Capitalizing on the a-priori distribution of channel coefficients, reference~\cite{Jiang2023EM} introduces a modified expectation-maximization approximate message passing (AMP) algorithm for CE, followed by AUD through the posterior support probabilities.
Reference~\cite{wang2022two} proposes a two-stage AUD and CE method, in which AUD is first conducted via adjacent APs utilizing vector AMP, and CE is performed through a linear estimator.
Reference~\cite{guo2022joint} performs joint AUD and CE through a single-measurement-vector-based minimum mean square error (MMSE) estimation approach at each AP independently.
Considering both centralized and edge computing paradigms, reference~\cite{Ke2021massive} presents an AMP-based approach for the joint AUD and CE while addressing quantization accuracy.
In millimeter-wave systems, reference~\cite{Johnston2022model} introduces two distinct algorithms for joint AUD and CE, leveraging the inherent characteristic that each UE's channel predominantly comprises a few significant propagation paths.
Reference~\cite{Di2022adaptive} presents a joint AUD and DD (JAD) algorithm, employing an adaptive AP selection method based on local log-likelihood ratios.
Reference~\cite{Femenias2023repetition} performs AUD, MMSE-based CE, and successive interference cancellation (SIC)-based data decoding under a probabilistic $K$-repitition scheme.
Reference~\cite{zhou2023active} first presents a joint AUD and CE approach for grant-free transmission using orthogonal time frequency space (OTFS) modulation in low Earth orbit (LEO) satellite communication systems. 
Subsequently, it introduces a least squares-based parallel time domain signal detection method.
Reference~\cite{Iimori2021grant} presents a Gaussian approximation-based Bayesian message passing algorithm for JACD, combined with an advanced low-coherence pilot design. 
Our previous work in~\cite{sun2023deep_unfolding} introduced a DU-based JAD algorithm, in which all algorithm hyper-parameters are optimized using machine learning. 
In addition, our study in~\cite{sun2023joint_arxiv} presents a box-constrained FBS algorithm designed for JACD.
Unlike previous methods, this paper tackles the task of JACD for massive grant-free transmission in cell-free systems. 
To improve JACD performance, we capture the sporadic UE activity more accurately by representing both the channel matrix and the data matrix as sparse matrices.

\subsubsection{Joint Active User Detection, Channel Estimation, and Data Detection for Single-Cell Massive Grant-Free Transmission}

JACD for single-cell massive grant-free transmission has been extensively investigated in~\cite{zou2020alow, Di2022joint, zhang2023joint, jiang2020joint, bai2023deep, Bian2023joint, shen2023joint}.
Considering low-precision data converters, reference~\cite{zou2020alow} utilizes bilinear generalized AMP (Bi-GAMP) with belief propagation algorithms for JACD in single-cell mMTC systems.
Reference~\cite{Di2022joint} proposes a bilinear message-scheduling generalized AMP for JACD, in which the channel decoder beliefs are used to refine AUD and DD.
Reference~\cite{zhang2023joint} develops a Bi-GAMP algorithm for JACD,  capturing the row-sparse channel matrix structure stemming from channel correlations.
Reference~\cite{jiang2020joint} divides the JACD scheme into slot-wise AUD and joint signal and channel estimation, which are addressed using message passing.
Reference~\cite{bai2023deep} combines AMP and belief propagation (BP) to perform JACD for asynchronous mMTC systems, in which the UEs transmit different lengths of data packets.
Reference~\cite{Bian2023joint} introduces a turbo-structured receiver for JACD and data decoding, utilizing the channel decoder's information to improve CE and DD performance.
Reference~\cite{shen2023joint} introduces a JACD algorithm based on message passing and Markov random fields for LEO satellite-enabled mMTC scenarios, which employs OTFS modulation to capitalize on the sparsity in the delay-Doppler-angle domain.
In contrast to these message-passing-based JACD methods that have been designed for single-cell systems that primarily focus on UE activity sparsity, we consider the JACD problem in cell-free systems by taking into account two distinct sources of sparsity in the channel matrix: (i) column sparsity, which stems from the sporadic UE activity in mMTC scenarios, and (ii) block sparsity within each non-zero column, which stems from the vast discrepancies in large-scale channel fading between UEs and distributed APs in cell-free systems~\cite{song2022joint}. 
Furthermore, we propose our JACD methods within the FBS framework, incorporating efficient strategies to improve the JACD performance and reduce computational complexity.

\subsubsection{Deep-Unfolding for  Massive Grant-Free Transmission and Cell-Free Systems}
DU techniques~\cite{Balatsoukas2019deep, hershey2014deep, Jagannath2021redefining, Ye2021deep} have increasingly found application in the domain of massive grant-free transmission and cell-free systems~\cite{Johnston2022model,bai2023deep, Bai2022prior, Ma2023model,shi2020sparse,gao2023hybrid, Liu2022model}, which adeptly utilize backpropagation and stochastic gradient descent to automatically learn algorithm hyper-parameters. 
Reference~\cite{Johnston2022model} employs DU to a linearized alternating direction method of multipliers and vector AMP, improving the performance of joint AUD and CE.
Reference~\cite{bai2023deep} applies DU to AMP-BP to improve JACD performance, fully exploiting the three-level sparsity inherent in UE activity, transmission delay, and data length diversity.
Reference~\cite{Bai2022prior} introduces a model-driven sparse recovery algorithm to estimate the sparse channel matrix in mMTC scenarios, effectively utilizing the a-priori knowledge of partially known supports.
Reference~\cite{Ma2023model} unfolds AMP, tailored for JAD in mMTC under single-phase non-coherent schemes, wherein the embedded parameters are trained to mitigate the performance degradation caused by the non-ideal i.i.d. model.
Reference~\cite{shi2020sparse} uses DU together with an iterative shrinkage thresholding algorithm for joint AUD and CE, wherein multiple computable matrices are treated as trainable parameters, thereby providing improving optimization flexibility.
Reference~\cite{gao2023hybrid} proposes a DU-based multi-user beamformer for cell-free systems, improving robustness to imperfect channel state information (CSI), where APs are equipped with fully digital or hybrid analog-digital arrays.
Reference~\cite{Liu2022model} unfolds a zero-forcing algorithm to achieve multi-user precoding, reducing complexity and improving the robustness under imperfect CSI.
In contrast to these results, we deploy DU to train all hyper-parameters in our proposed algorithms to improve the JACD performance and employ approximations for high-complexity steps to decrease computational complexity.
Furthermore, we train the hyper-parameters of our soft-output AUD module that generates information on UE activity probability.

\subsection{Notation}

Matrices, column vectors, and sets are denoted by uppercase boldface letters, lowercase boldface letters, and uppercase calligraphic letters, respectively. 
$\mathbf{A}(m,n)$ stands for the element of matrix $\mathbf{A}$ at the $m$th row and $n$th column, and $\mathbf{a}(m)$ stands for the $m$th entry of the vector $\mathbf{a}$.
The $M\times N$ all-ones matrix is $\mathbf{1}_{M\times N}$ and the $M\times M$ identity matrix is $\mathbf{I}_M$. 
The unit vector, which is zero except for the $m$th entry, is~$\mathbf{e}_m$, and the zero vector is $\mathbf{0}$; the dimensions of these vectors will be clear from the context. 
We use hat symbols to refer to the estimated values of a variable, vector, or matrix.
The superscripts $\left(\cdot\right)^T$ and $\left(\cdot\right)^H$ represent transpose and conjugate transpose, respectively, and the superscript $(k)$ denotes the $k$th iteration.
The Frobenius norm is denoted by $\|\cdot\|_F$. 
In addition, $\operatorname{diag}\{x_1,\ldots,x_N\}$ stands for a diagonal matrix with entries $x_1,\ldots,x_N$ on the main diagonal; $\operatorname{det}(\mathbf{A})$ stands for the determinant of $\mathbf{A}$. 
The cardinality of a set $\mathcal{Q}$ is $|\mathcal{Q}|$. 
The operators $\odot$ and $\propto$ denote Hadamard product and proportional relationships.
For $\mathbf{x} \in \mathbb{C}^N$, $\text{Re}\{\mathbf{x}\}\in \mathbb{R}^N$ and $\text{Im}\{\mathbf{x}\}\in \mathbb{R}^N$ represent its real and imaginary part, respectively. 
$\mathbb{P}\{\cdot\}$ and $\mathbb{E}\{\cdot\}$ denote probability and expectation, respectively; the indicator function $\mathbb{I}\left\{\cdot\right\}$ returns $1$ for valid conditions and $0$ otherwise. 
The multivariate complex Gaussian probability distribution with mean vector $\mathbf{m}$ and covariance matrix $\boldsymbol{\Sigma}$ evaluated at $\mathbf{x}\in\mathbb{C}^M$~is 
\begin{align}
\mathcal{CN}\left(\mathbf{x};\mathbf{m},\boldsymbol{\Sigma}\right)\triangleq \frac{\exp\left(-\left(\mathbf{x}-\mathbf{m}\right)^H\boldsymbol{\Sigma}^{-1}\left(\mathbf{x}-\mathbf{m}\right)\right)}{\pi^M \operatorname{det}\left(\boldsymbol{\Sigma}\right)},
\end{align} 
and the symbol $\triangleq$ means ``is defined as.'' Besides, the shrinkage operation is defined as
\begin{equation}
	\label{shrinkage_definition}
	\begin{aligned}
		\textsf{Shrinkage}\left(\mathbf{x},\mu\right) \triangleq  \left\{\begin{array}{l}
			\mathbf{x}\frac{\max\{\left\|\mathbf{x}\right\|_F-\mu,0\}}{\left\|\mathbf{x}\right\|_F}, \text{ if }\mathbf{x}\neq \mathbf{0},\\
		\mathbf{0}, \text{ if }\mathbf{x}=\mathbf{0}.
		\end{array}\right.
	\end{aligned}
\end{equation}
and the element-wise clamp function is defined as
\begin{equation}
	\label{clamp_definition}
    \begin{aligned}
        \textsf{Clamp}\left(\mathbf{x},L,U\right)\triangleq &\min\left\{\max\left\{\text{Re}\left\{\mathbf{x}\right\},L\right\},U\right\} \\
        &+ j \min\left\{\max\left\{\text{Im}\left\{\mathbf{x}\right\},L\right\},U\right\}.
    \end{aligned}
\end{equation}

\subsection{Paper Outline}
The rest of this paper is organized as follows.
Section~\ref{Section_II} presents the necessary prerequisites for the subsequent derivations.
Section~\ref{Section_III} introduces the system model and formulates the JACD optimization problem. 
Section~\ref{Section_IV} introduces two distinct JACD algorithms: (i) box-constrained FBS and (ii) PME-based JACD. 
Section~\ref{Section_V} deploys DU techniques and details the AUD module. 
Section~\ref{Section_VI} investigates the efficacy of the proposed algorithms via Monte--Carlo simulations. 
Section~\ref{Section_VII} concludes the paper. Proofs are relegated to the~appendices. 

\section{Prerequisites}\label{Section_II}
In this section, we introduce some background knowledge and mathematical definitions related to the subsequent sections. 
Specifically, the introduction to FBS will provide insights into the derivation of the JACD algorithms under the FBS framework in Section~\ref{Section_IV}. 
In addition, Definitions 1 and 3 are primarily used for the derivation of the PME-based JACD algorithm in Section~\ref{Section_IV_B}, and Definition 2 is mainly utilized in the derivation of the DU-POEM algorithm in Section~\ref{Section_V_B}.

\subsection{A Brief Introduction to FBS}\label{Section_II_A}
FBS, also known as proximal gradient methods, is widely used for solving a wide variety of convex optimization problems~\cite{goldstein2014field,beck2009fast}. 
FBS splits the objective function into two components: a smooth function, denoted as $f(\mathbf{S})$, and another not necessarily smooth function, $g(\mathbf{S})$, and solves the following optimization problem:
\begin{equation}
	\begin{aligned}
		\hat{\mathbf{S}} = {\arg\min}_{\mathbf{S}}  \;f(\mathbf{S})
		+ g(\mathbf{S}).
	\end{aligned}
\end{equation}
FBS iteratively performs a gradient step in the smooth function $f(\mathbf{S})$ (denoted as the forward step) and a proximal operation to find a solution in the vicinity of the minimizer of the function $g(\mathbf{S})$ (denoted as the backward step). 
The forward step proceeds~as 
\begin{equation}
\hat{\mathbf{S}}^{(k)}=\mathbf{S}^{(k)}-\tau^{(k)} \nabla f \!\left(\mathbf{S}^{(k)}\right)\!,
 \label{forward}
\end{equation}
where the superscript $(k)$ denotes the $k$th iteration, $\tau^{(k)}$ represents the step size at iteration $k$, and $\nabla f \left(\mathbf{S}\right)$ is the gradient of $f(\mathbf{S})$. The backward step proceeds as
\begin{equation}
\mathbf{S}^{(k+1)}={\arg\min}_{\mathbf{S}}  \;\frac{1}{2} \left\|\mathbf{S}-\hat{\mathbf{S}}^{(k)}\right\|_F^2 
		+ \tau^{(k)} g(\mathbf{S}).
\end{equation}
This process is iterated for $k=1,2,\ldots$ until a predefined convergence criterion is met or a maximum number of $K$  iterations has been reached. 

\subsection{Some Mathematical Definitions}\label{Section_II_B}
Here, we introduce Definitions 1 and 2 to specify the probability distributions that the random vector $\mathbf{x}$ may follow.

\textit{\textbf{Definition 1:} For a random vector $\mathbf{x}$ taken from the discrete set $\mathcal{S}$ ($\mathbf{0}\in\mathcal{S}$), we call 
$\mathbf{x}$ follows the $\theta$-mixed discrete uniform distribution on $\mathcal{S}$, denoted as $\mathbf{x}\sim U_{\theta,\mathcal{S}}\{\mathbf{x}\}$, if $\mathbb{P}\{\mathbf{x}\}=U_{\theta,\mathcal{S}}\{\mathbf{x}\}=\frac{\theta}{|\mathcal{S}|-1}\mathbb{I}\{\mathbf{x}\ne \mathbf{0}\} + (1-\theta)\mathbb{I}\{\mathbf{x}= \mathbf{0}\},\forall \mathbf{x}\in \mathcal{S}$, where $\theta\in (0,1)$ is the probability of $\mathbf{x}$ being non-zero.}

\textit{\textbf{Definition 2:} For a random vector $\mathbf{x}$ in a discrete set $\mathcal{S}$ ($\mathbf{0}\notin\mathcal{S}$), we call 
$\mathbf{x}$ follows a discrete uniform distribution on~$\mathcal{S}$, denoted as $\mathbf{x}\sim U_\mathcal{S}\{\mathbf{x}\}$,  if $\mathbb{P}\{\mathbf{x}\}=U_\mathcal{S}\{\mathbf{x}\}=\frac{1}{|\mathcal{S}|}, \forall \mathbf{x}\in\mathcal{S}$.}

We also define the PME of a vector $\mathbf{x}$ under the specific conditions as follows:

\textit{\textbf{Definition 3:} Given the observation vector $\hat{\mathbf{x}}=\mathbf{x} +\mathbf{e}$ with $\mathbf{x}\in\mathcal{S}$ following the prior distribution $\mathbb{P}\{\mathbf{x}\}$ and Gaussian estimation error $\mathbf{e}\sim\mathcal{CN}\left(\mathbf{0},N_{\text{e}}\mathbf{I}\right)$, we call
\begin{align}
 \text{PME}(\hat{\mathbf{x}},\mathcal{S},\mathbb{P}\{\mathbf{x}\},N_{\text{e}})=\frac{\int_{\mathbf{x}\in\mathcal{S}}\mathbb{P}\{\mathbf{x}\}\mathcal{CN}(\mathbf{x};\hat{\mathbf{x}},N_{\text{e}}\mathbf{I})\mathbf{x}\,\mathrm{d}\mathbf{x}}{\int_{\mathbf{x}\in\mathcal{S}}\mathbb{P}\{\mathbf{x}\}\mathcal{CN}(\mathbf{x};\hat{\mathbf{x}},N_{\text{e}}\mathbf{I})\,\mathrm{d}\mathbf{x}},
\end{align}
the PME of $\mathbf{x}$ under the prior $\mathbb{P}\{\mathbf{x}\}$.}

\section{System Model and Problem Formulation} \label{Section_III}
\begin{figure}[tp]
	\centering
	\includegraphics[width=0.9\columnwidth]{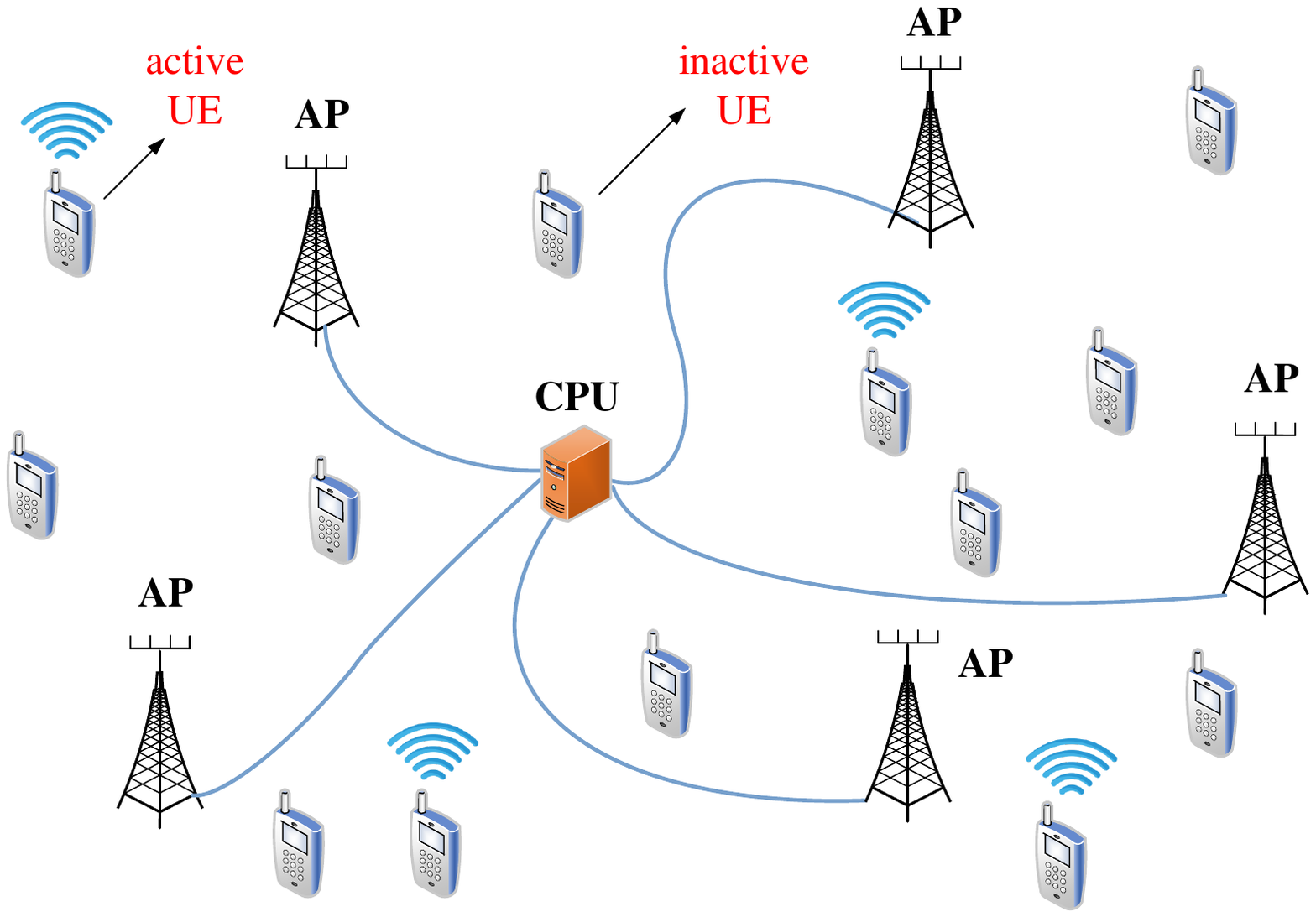}
	\vspace{-0.2cm}
	\caption{Illustration of mMTC in a cell-free wireless communication system.}
	\label{scenarios}
\end{figure}

We consider an mMTC scenario in a cell-free wireless communication system as illustrated in Fig.~\ref{scenarios}.
We model the situation using $P$ distributed APs with $M$ antennas that serve $N$ single-antenna sporadically active UEs.
We assume that the $N_a\ll N$ active UEs are synchronized in time and simultaneously transmit uplink signals to APs over $R$ resource elements, assuming frequency-flat and block-fading~channels.

\subsection{System Model}
Following our previous work in~\cite{sun2023joint_arxiv,sun2023deep_unfolding}, we model the input-output relation of this scenario as follows:
\begin{equation}
	 \mathbf{Y} = \sum_{n=1}^{N}\xi_n \mathbf{h}_{n}\bar{\mathbf{x}}_n^T + \mathbf{N}.
	\label{system_model}
\end{equation}
Here, the received signal matrix from all APs is denoted by~$\mathbf{Y}\in\mathbb{C}^{MP\times R}$,  and $\xi_n\in\{0,1\}$ is $n$th UE's activity indicator with $\xi_n=1$ indicating the $n$th UE is active and $\xi_n=0$ otherwise. 
We assume that the activity between UEs is independent and all UEs have the same activity probability $\alpha$, i.e., $\mathbb{P}\{\xi_n=1\}=\alpha, \forall n$. 
The channel vector between the $n$th UE and all APs is $\mathbf{h}_{n}\triangleq[\mathbf{h}_{n,1}^T,\ldots,\mathbf{h}_{n,P}^T]^T\in\mathbb{C}^{MP}$ with $\mathbf{h}_{n,p}\in\mathbb{C}^{M}$ being the channel vector between the $n$th UE and the $p$th AP.
The $n$th UE's signal vector $\bar{\mathbf{x}}_n\triangleq[\mathbf{x}_{\text{P},n}^T,\bar{\mathbf{x}}_{\text{D},n}^T]^T\in\mathbb{C}^{R}$ consists of the pilot vector $\mathbf{x}_{\text{P},n}\in\mathbb{C}^{R_{\text{P}}}$ and the data vector $\bar{\mathbf{x}}_{\text{D},n}\in\mathcal{Q}^{R_{\text{D}}}$ with data entries independently and uniformly sampled from the constellation set $\mathcal{Q}$. 
Entries in the noise matrix~$\mathbf{N}\in\mathbb{C}^{MP\times R}$ are assumed to be i.i.d. circularly-symmetric complex Gaussian variables with variance $N_0=1$.

\begin{figure}[tp]
	\centering
	\includegraphics[width=0.45\textwidth]{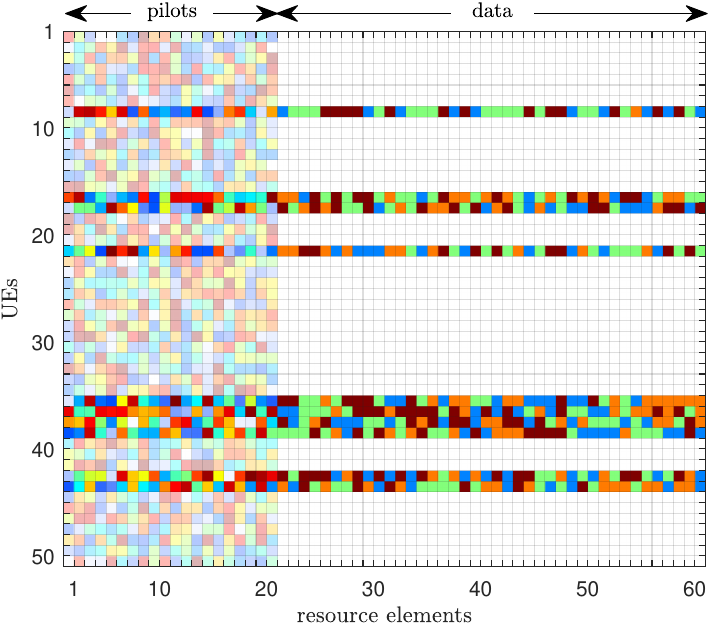}
	\caption{Illustration of the signal matrix $\mathbf{X}$, where only 10 out of 50 UEs are active. Active UEs transmit unique pilots of length $R_{\text{P}} = 20$ and data symbols of length $R_{\text{D}} = 40$, represented by dark-colored squares. Inactive UEs transmit no signals; however, their unique pilots, which are known to the BS and represented by light-colored squares, can be used to improve JACD~performance.}
	\label{X}
\end{figure}

\newcounter{TempEqCnt}
\setcounter{TempEqCnt}{\value{equation}}
\setcounter{equation}{13}
\begin{figure*}
\vspace{-1ex}
\begin{equation}
		\mathcal{P}_1:\;\left\{ {\hat{\mathbf{H}},{{\hat {\mathbf{X}}}_{\text{D}}}} \right\} =\mathop {\arg\min}_{\scriptstyle{\mathbf{H}\in\mathbb{C}^{MP\times N}}\hfill\atop
			{	\scriptstyle{\mathbf{x}_{\text{D},n}\in\bar{\mathcal{Q}}, \forall n}\hfill}} \frac{1}{2}\left\|\mathbf{Y}-\mathbf{H}\left[\mathbf{X}_{\text{P}},\mathbf{X}_{\text{D}}\right]\right\|_F^2 
		+\mu_h\sum_{n=1}^N\sum_{p=1}^P\left\|\mathbf{h}_{n,p}\right\|_F+ \mu_x\sum_{n=1}^N\left\|\mathbf{x}_{\text{D},n}\right\|_F.
	\label{P1}
\end{equation}
\hrulefill
\end{figure*}

\setcounter{equation}{15}
\begin{figure*}
\vspace{-1ex}
\begin{equation}
		\mathcal{P}_2:\left\{ {\hat{\mathbf{H}},{{\hat {\mathbf{X}}}_{\text{D}}}} \right\} =\mathop {\arg\min}_{\scriptstyle{\mathbf{H}\in\mathbb{C}^{MP\times N}}\hfill\atop
			{	\scriptstyle{{\mathbf{X}}_{\text{D}}\in\mathcal{B}^{N\times R_{\text{D}}}}\hfill}}  \frac{1}{2}\left\|\mathbf{Y}-\mathbf{H}\left[\mathbf{X}_{\text{P}},\mathbf{X}_{\text{D}}\right]\right\|_F^2   + \mu_h\sum_{n=1}^N\sum_{p=1}^P\left\|\mathbf{h}_{n,p}\right\|_F
		+ \mu_x\sum_{n=1}^N\left\|\mathbf{x}_{\text{D},n}\right\|_F+ \lambda\, \mathcal{C}\left(\mathbf{X}_{\text{D}}\right).
	\label{P2}
\end{equation}
\hrulefill
\vspace{-3ex}
\end{figure*}
\setcounter{equation}{\value{TempEqCnt}}

For ease of notation, we rewrite the system model in \eqref{system_model} as 
\begin{align}
	\textstyle
	\mathbf{Y} 
 = \mathbf{H}\left[\mathbf{X}_{\text{P}},\mathbf{X}_{\text{D}}\right]+\mathbf{N}
 = \mathbf{H}\mathbf{X}+\mathbf{N},
	\label{system_model2}
\end{align}
where $\mathbf{Y}=[\mathbf{Y}_{\text{P}},\mathbf{Y}_{\text{D}}]$ with $\mathbf{Y}_{\text{P}}\in\mathbb{C}^{MP\times R_{\text{P}}}$ and $\mathbf{Y}_{\text{D}}\in\mathbb{C}^{MP\times R_{\text{D}}}$ being the received pilot matrix and received data matrix, respectively. 
The channel matrix is given by $\mathbf{H}\triangleq \left[\xi_1\mathbf{h}_{1}, \ldots, \xi_N\mathbf{h}_{N}\right] \in\mathbb{C}^{MP\times N}$. 
The signal matrix $\mathbf{X}=[\mathbf{X}_{\text{P}},\mathbf{X}_{\text{D}}]\in\mathbb{C}^{N\times R}$ contains the pilot matrix $\mathbf{X}_{\text{P}} \triangleq [\mathbf{x}_{\text{P},1},\ldots,\mathbf{x}_{\text{P},N}]^T\in\mathbb{C}^{N\times R_{\text{P}}}$ and the data matrix $\mathbf{X}_{\text{D}} \triangleq [\mathbf{x}_{\text{D},1},\ldots, \mathbf{x}_{\text{D},N}]^T\in\mathbb{C}^{N\times R_{\text{D}}}$, where  $\mathbf{x}_{\text{D},n}\triangleq\xi_n\bar{\mathbf{x}}_{\text{D},n}\in \bar{\mathcal{Q}}$ with $\bar{\mathcal{Q}}\triangleq \{\mathcal{Q}^{R_{\text{D}}},\mathbf{0}\}$.

\begin{figure}[tp]
	\centering
    \includegraphics[width=0.45\textwidth]{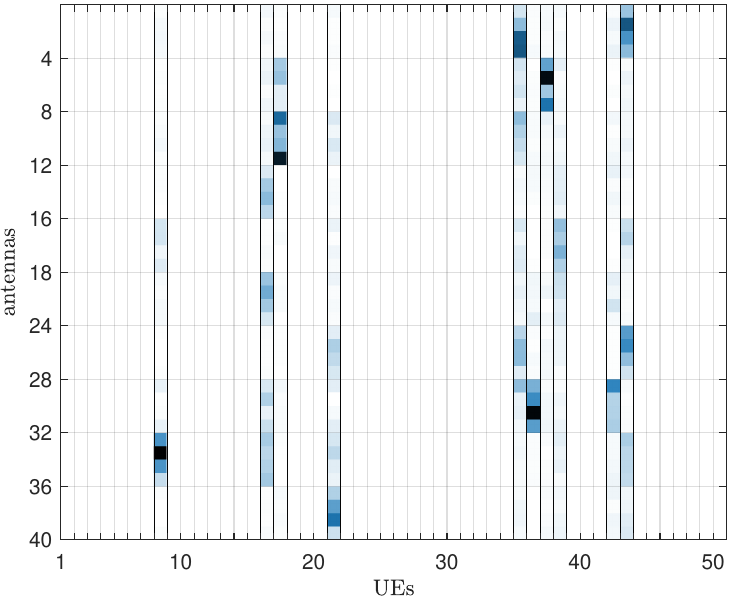}
	\caption{Illustration of a channel matrix $\mathbf{H}$ for $10$ APs with $4$ antennas each and $50$ UEs, where $10$ UEs are active. Darker colors indicate larger absolute values; the boxed columns correspond to the active UEs (cf. Fig.~\ref{X}).}
	\label{channel}
\end{figure}

A typical signal matrix $\mathbf{X}$ is depicted in Fig.~\ref{X}.  
Here, the data matrix $\mathbf{X}_{\text{D}}$ is a row-sparse matrix due to sporadic UE activity, and the pilot matrix $\mathbf{X}_{\text{P}}$ is characterized as a non-sparse matrix, retaining all known pilot information for subsequent optimization.  
A typical channel matrix~$\mathbf{H}$ is depicted in Fig.~\ref{channel}, where sparsity is due to two reasons: (i) the UEs' sporadic activity results in column sparsity and (ii) the inherent discrepancies in large-scale fading between UEs and different APs, caused by their varying distances, lead to the block sparsity within each non-zero column.

While row/column sparsity and block sparsity have been widely studied in compressed sensing, the intricate interplay of these sparsities in our setting remains largely unexplored. 
In our system model \eqref{system_model2}, the estimation of both $\mathbf{H}$ and $\mathbf{X}$ poses a significant challenge, as $\mathbf{H}$ exhibits both column and row sparsity, while $\mathbf{X}_{\text{D}}$ displays row sparsity. 
In the subsequent problem formulation and algorithm derivation, we will effectively leverage the sparsity of both the channel matrix $\mathbf{H}$ and the data matrix $\mathbf{X}_{\text{D}}$ to enhance JACD performance.

\subsection{Problem Formulation}
Using the system model~\eqref{system_model2}, we formulate the maximum-a-posteriori JACD problem for massive grant-free transmission in cell-free wireless communication systems as~\cite{sun2023joint_arxiv,sun2023deep_unfolding}
\begin{equation}
\left\{ \hat{\mathbf{H}},{{\hat {\mathbf{X}}}_{\text{D}}} \right\} =\mathop {\arg \max }\limits_{\scriptstyle{\mathbf{H}\in\mathbb{C}^{MP\times N}}\hfill\atop
			{	\scriptstyle{\mathbf{x}_{\text{D},n}\in\bar{\mathcal{Q}}, \forall n}\hfill}} P\!\left(\mathbf{Y}|\mathbf{H},\mathbf{X}_{\text{D}}\right)\!P\!\left(\mathbf{H}\right)\!P\!\left(\mathbf{X}_{\text{D}}\right),
	\label{joint_problem}
\end{equation}
where the channel law $P\!\left(\mathbf{Y}|\mathbf{H},\mathbf{X}_{\text{D}}\right)$ is given by
\begin{equation} 
P\!\left(\mathbf{Y}|\mathbf{H},\mathbf{X}_{\text{D}}\right) \propto\exp\Big(-\frac{\left\|\mathbf{Y}-\mathbf{H}\left[\mathbf{X}_{\text{P}},\mathbf{X}_{\text{D}}\right]\right\|_F^2}{N_0}\Big).
	\label{P_Y_given_H_X_D}
\end{equation}
Here, we employ the complex-valued block-Laplace model and Laplace model for block sparsity in $\mathbf{H}$ and column sparsity in $\mathbf{X}_{\text{D}}$, respectively, as follows~\cite{sun2023joint_arxiv,sun2023deep_unfolding}
\begin{align}
& P\!\left(\mathbf{H}\right) \propto \prod_{n=1}^{N}\prod_{p=1}^{P}\exp\!\left(-2\mu_h\left\|\mathbf{h}_{n,p}\right\|_F\right)\!,\label{P_H}\\
& P\!\left(\mathbf{X}_{\text{D}}\right) \propto \prod_{n=1}^{N}\exp\!\left(-2\mu_x\left\|\mathbf{x}_{\text{D},n}\right\|_F\right)\!, \label{P_D}
\end{align}
with $\mu_h$ and $\mu_x$ indicating the sparsity levels of $\mathbf{H}$ and $\mathbf{X}_{\text{D}}$,~respectively.

By inserting \eqref{P_Y_given_H_X_D}, \eqref{P_H}, and \eqref{P_D} into \eqref{joint_problem}, we can rewrite the JACD problem as problem $\mathcal{P}_1$ with $N_0=1$, expressed in problem \eqref{P1} above. 
This problem aims to estimate the channel matrix $\mathbf{H}$ and the data matrix $\mathbf{X}_{\text{D}}$ using the received signal matrix $\mathbf{Y}$ and the pilot matrix $\mathbf{X}_{\text{P}}$, in which UE activity is indicated by the column sparsity of $\mathbf{H}$ and row sparsity of~$\mathbf{X}_{\text{D}}$.

\subsection{Problem Relaxation}
The discrete set $\bar{\mathcal{Q}}$ renders $\mathcal{P}_1$ a discrete-valued optimization~problem for which a na\"ive exhaustive search is infeasible. To circumvent this limitation, as in~\cite{sun2023joint_arxiv, sun2023deep_unfolding}, we relax the discrete set $\bar{\mathcal{Q}}$ to its convex hull $\mathcal{B}^{R_{\text{D}}}$, thereby transforming $\mathcal{P}_1$ into a continuous-valued optimization problem. 
The set $\mathcal{B}$ is given~by\footnote{For quadrature phase shift keying (QPSK), the set $\mathcal{B}$ can be specified as  $\mathcal{B}=\{x\in\mathbb{C}:-B\le\text{Re}\{x\}\le B,-B\le\text{Im}\{x\}\le B\}$.}
\setcounter{equation}{14}
\begin{equation}
 \textstyle \mathcal{B}\triangleq\left\{\sum_{i=1}^{|\mathcal{Q}|}\delta_i q_i:\,q_i\in\mathcal{Q},\,\delta_i\ge 0,\forall i;\,\sum_{i=1}^{|\mathcal{Q}|}\delta_i=1\right\}\!.
\end{equation}

To push the entries in the data matrix $\mathbf{X}_{\text{D}}$ towards points in discrete set $\bar{\mathcal{Q}}$, two distinct methodologies can be used: \textit{Method (i)} introduces a regularizer (also called penalty term) into the objective function of $\mathcal{P}_1$~\cite{sun2023joint_arxiv, Oscar2017bit} and \textit{method~(ii)} leverages the PME to denoise the estimated data matrix $\mathbf{X}_{\text{D}}$~\cite{sun2023deep_unfolding}.
As such, we can transform the original discrete-valued optimization problem~$\mathcal{P}_1$ into problem $\mathcal{P}_2$ in \eqref{P2} above. 
The penalty parameter $\lambda$ is  $\lambda> 0$ for \textit{method (i)} and $\lambda= 0$ for \textit{method (ii)}. 
For the regularizer $\mathcal{C}\left(\mathbf{X}_{\text{D}}\right)$, many alternatives are possible, such as $\mathcal{C}\left(\mathbf{X}_{\text{D}}\right)=-\|\mathbf{X}_{\text{D}}\odot\mathbf{X}_{\text{D}}^*-B^2\mathbf{1}_{N\times R_{\text{D}}}\|_F^2$ utilized in~\cite{sun2023joint_arxiv}. 
In the next section, we will develop the FBS algorithm based on \textit{method (i)} and \textit{method (ii)} respectively to efficiently compute approximate solutions to problem $\mathcal{P}_2$.

\section{JACD Algorithms}\label{Section_IV}
We develop two JACD algorithms that utilize FBS, each leveraging specific techniques to improve JACD performance. 
Specifically, the first algorithm based on \textit{method~(i)} is called the box-constrained FBS algorithm~\cite{sun2023joint_arxiv}, which utilizes a regularizer within the objective function to guide the estimated symbols toward the discrete constellation points, thereby improving DD accuracy.
The second algorithm based on \textit{method~(ii)} is called the PME-based JACD algorithm, which employs PME to modify estimated data to further improve DD performance.

\subsection{Box-Constrained FBS Algorithm for JACD}\label{III_B}
For this method, we utilize the FBS with the incorporation of the regularizer $\mathcal{C}(\mathbf{X}_{\text{D}})$ to approximately solve the non-convex problem $\mathcal{P}_2$. 
Using the  definition $\mathbf{S}\triangleq[\mathbf{H}^H, \mathbf{X}_{\text{D}}]^H\in\mathbb{C}^{(MP+R_{\text{D}})\times N}$ in~\cite{sun2023joint_arxiv, sun2023deep_unfolding, song2022joint}, we can split the objective function of $\mathcal{P}_2$ into the two functions 
\setcounter{equation}{16}
\begin{align}
\!\!\!f(\mathbf{S})&= \frac{1}{2}\left\|\mathbf{Y}-\mathbf{H}\left[\mathbf{X}_{\text{P}},\mathbf{X}_{\text{D}}\right]\right\|_F^2  \!+\! \lambda\,\mathcal{C}(\mathbf{X}_{\text{D}}),\label{fS}\\
\!\!\!g(\mathbf{S})&= \mu_h\!\!\sum_{n=1}^N\sum_{p=1}^P\left\|\mathbf{h}_{n,p}\right\|_F \!+\!\mu_x\!\!\sum_{n=1}^N\left\|\mathbf{x}_{\text{D},n}\right\|_F \!+\! \mathcal{X}\!\left(\mathbf{X}_{\text{D}}\right),
 \label{fS_gS}
\end{align}
where $\mathcal{X}\left(\mathbf{X}_{\text{D}}\right)=+\infty\cdot\mathbb{I}\left\{\mathbf{X}_{\text{D}}\notin\mathcal{B}^{N\times R_{\text{D}}}\right\}$ represents the constraint ${\mathbf{X}}_{\text{D}}\in\mathcal{B}^{N\times R_{\text{D}}}$ in problem $\mathcal{P}_2$. 
Following the FBS framework outlined in Section~\ref{Section_II_A}, the corresponding forward and backward steps can be specified as follows.

\subsubsection{Forward Step}
Given the expression of $f(\mathbf{S})$ in \eqref{fS}, we can derive the gradient of $f \!\left(\mathbf{S}\right)\!$ with respect to $\mathbf{S}$ in the forward step \eqref{forward} as follows~\cite{sun2023joint_arxiv}:
\begin{equation}
    \nabla f (\mathbf{S}) =\left[\left(\frac{\partial f }{\partial \mathbf{H}^*} \right)^T,\left(\frac{\partial f }{\partial \mathbf{X}_{\text{D}}^T} \right)^T\right]^T,
    \label{gradient}
\end{equation} 
where 
\begin{align}
\frac{\partial f }{\partial \mathbf{H}^*}& =-\!\left(\mathbf{Y}-\mathbf{H}\mathbf{X}\right)\!\mathbf{X}^H,\\
\frac{\partial f }{\partial \mathbf{X}_{\text{D}}^T}& =-\!\left(\mathbf{Y}_{\text{D}}-\mathbf{H}\mathbf{X}_{\text{D}}\right)\!^H\mathbf{H} + \lambda\,\frac{\partial \mathcal{C}\!\left(\mathbf{X}_{\text{D}}\right)\!}{\partial\mathbf{X}_{\text{D}}^T}.
\end{align}
The expression $ \frac{\partial \mathcal{C}\!\left(\mathbf{X}_{\text{D}}\right)\!}{\partial \mathbf{X}_{\text{D}}^T} $ varies for various choices of $ \mathcal{C}\left(\mathbf{X}_{\text{D}}\right)$. 
For instance, $\frac{\partial \mathcal{C}\!\left(\mathbf{X}_{\text{D}}\right)\!}{\partial \mathbf{X}_{\text{D}}^T}=-4 \left(\mathbf{X}_{\text{D}}^*\odot\left(\mathbf{X}_{\text{D}}\odot\mathbf{X}_{\text{D}}^*-B^2\mathbf{1}_{N\times R_{\text{D}}}\right)\right)^T$ for $\mathcal{C}\left(\mathbf{X}_{\text{D}}\right)=-\left\|\mathbf{X}_{\text{D}}\odot\mathbf{X}_{\text{D}}^*-B^2\mathbf{1}_{N\times R_{\text{D}}}\right\|_F^2$.

\subsubsection{Backward Step} 
The proximal operator for $\mathbf{S}$ can be decomposed into separate proximal operators for $\mathbf{H}$ and $\mathbf{X}_{\text{D}}$, respectively. 
The proximal operator for $\mathbf{H}$ is 
\begin{equation}
\begin{aligned}
\mathbf{H}^{(k+1)}\!=\!\!\mathop {\arg\min}_{\scriptstyle{\mathbf{H}\in\mathbb{C}^{MP\times N}}} \!\frac{1}{2} \left\|\mathbf{H}\!-\!\hat{\mathbf{H}}^{(k)}\right\|_F^2\!\! + \!\tau^{(k)}\mu_h\!\!\sum_{n=1}^N\sum_{p=1}^P\left\|\mathbf{h}_{n,p}\right\|_F\!,
\end{aligned}
\label{backward_H}
\end{equation}
with $\tau^{(k)}$ representing the step size at iteration $k$ in the forward step \eqref{forward}. 
The closed-form solution of problem \eqref{backward_H} is given by~\cite{song2022joint,goldstein2014field,beck2009fast}
\begin{equation}
	\mathbf{h}_{n,p}^{(k+1)} = \textsf{Shrinkage}\left(\hat{\mathbf{h}}_{n,p}^{(k)},\tau^{(k)}\mu_h\right),
	\label{Shrinkage}
\end{equation}
with the shrinkage operation defined in \eqref{shrinkage_definition}.
The proximal operator for $\mathbf{X}_{\text{D}}$ can be decomposed as independent proximal operators for $\mathbf{x}_{\text{D},n},\,\forall n$ as follows  
\begin{align} \!\mathbf{x}_{\text{D},n}^{(k+1)}\!=\!\!\mathop{\arg\min}_{\scriptstyle{\mathbf{x}_{\text{D},n}\in\mathcal{B}^{R_{\text{D}}}}}   \frac{1}{2}\! \left\|\mathbf{x}_{\text{D},n}\!-\!\hat{\mathbf{x}}_{\text{D},n}^{(k)}\right\|_F^2 \! + \! \tau^{(k)}\mu_x\left\|\mathbf{x}_{\text{D},n}\right\|_F,
\label{xd_prox}
\end{align}
which is a convex optimization problem, and the optimal solution of \eqref{xd_prox} can be obtained by the KKT conditions outlined in~\cite{sun2023joint_arxiv}. 
For completeness, the following proposition details the closed-form solution to \eqref{xd_prox} and its proof is given in Appendix~\ref{proof of Proposition 1}.

\textit{\textbf{Proposition 1:} The optimal solution of \eqref{xd_prox} is given by $\mathbf{x}_{\text{D},n}^{(k+1)}=\left[\mathbf{I}_{R_{\text{D}}}, i\mathbf{I}_{R_{\text{D}}}\right]\mathbf{r}_n^{(k+1)}$, where $\mathbf{r}_n^{(k+1)}\in\mathbb{R}^{2R_{\text{D}}}$ is expressed~as:
\begin{equation}
\begin{aligned}
    \mathbf{r}_n^{(k+1)}(d) = \, &b\,\hat{\mathbf{r}}_n^{(k)}(d)\,\mathbb{I}\left\{d\notin\mathcal{S}_p\cup\mathcal{S}_q\right\}+B\,\mathbb{I}\left\{d\in\mathcal{S}_p, b\ne 0\right\}\\
    &-B\,\mathbb{I}\left\{d\in\mathcal{S}_q, b\ne 0\right\}.
\end{aligned}
\label{KKT_solution}
\end{equation}
Here, $\hat{\mathbf{r}}_n^{(k)} = [\text{Re}\{\hat{\mathbf{x}}_{\text{D},n}^{(k)}\}^T,\text{Im}\{\hat{\mathbf{x}}_{\text{D},n}^{(k)}\}^T]^T\in\mathbb{R}^{2R_{\text{D}} }$,  $\mathcal{S}_p\triangleq \{d:\,\mathbf{r}_{n,\text{tmp}}^{(k+1)}(d)>B\}$, and $\mathcal{S}_q\triangleq \{d:\,\mathbf{r}_{n,\text{tmp}}^{(k+1)}(d)<-B\}$, where $\textstyle\mathbf{r}_{n,\text{tmp}}^{(k+1)} =\textsf{Shrinkage}\big(\hat{\mathbf{r}}_n^{(k)},\tau^{(k)}\mu_x\big)$. 
In addition, $b$ is the solution of the quartic equation $2\sum_{d\notin \mathcal{S}_p\cup\mathcal{S}_q}\hat{\mathbf{r}}_n^{(k)}(d)^2 b^4 -4\sum_{d\notin \mathcal{S}_p\cup\mathcal{S}_q}\hat{\mathbf{r}}_n^{(k)}(d)^2 b^3 + (2\sum_{d\notin \mathcal{S}_p\cup\mathcal{S}_q}\hat{\mathbf{r}}_n^{(k)}(d)^2+|\mathcal{S}_p\cup\mathcal{S}_q|-2(\tau^d\mu_x)^2)b^2 -2|\mathcal{S}_p\cup\mathcal{S}_q| b + |\mathcal{S}_p\cup\mathcal{S}_q| = 0$ within the interval (0,1], if it exists; otherwise, $ b= 0 $.}

The pseudocode for the proposed box-constrained FBS algorithm is presented in Algorithm~\ref{alg1}.

\begin{algorithm}[t]
	\SetAlgoNoLine
	\caption{Box-Constrained FBS Algorithm}
	\label{alg1}
	\textbf{Input}: $\mathbf{Y}$, $\mathbf{X}_{\text{P}}$, $\mu_h$, $\mu_x$, $\lambda$, and $K$.\\
 \textbf{Initialization}: $\mathbf{S}^{(1)}=[(\mathbf{H}^{(1)})^H, \mathbf{X}_{\text{D}}^{(1)}]^H$.\\
 \For{$k=1,\ldots,K$}{
 \textit{Forward Step:} Calculate $\hat{\mathbf{S}}^{(k)}$ via \eqref{forward} and \eqref{gradient}.\\
 \textit{Backward Step:} Calculate $\mathbf{H}^{(k+1)}$ and $\mathbf{X}_{\text{D}}^{(k+1)}$ via \eqref{Shrinkage} and Proposition~1, respectively.
 }
	\textbf{Output}: $\mathbf{H}^{(k+1)}$ and $\mathbf{X}_{\text{D}}^{(k+1)}$.
\end{algorithm}

\setcounter{TempEqCnt}{\value{equation}}
\setcounter{equation}{26}
\begin{figure*}
\vspace{-1ex}
\begin{equation}
 \mathbf{x}_{\text{D},n}^{(k+1)} 
 =  \text{PME}\left(\hat{\mathbf{x}}_{\text{D},n}^{(k)},\bar{\mathcal{Q}},U_{\alpha,\bar{\mathcal{Q}}}\{\mathbf{x}\},N_{\text{e}}^{(k)}\right)
 =\frac{\sum_{\mathbf{x}\in\mathcal{Q}^{R_{\text{D}}}}{\frac{\alpha}{|\mathcal{Q}^{R_{\text{D}}}|}\,\mathcal{CN}\left(\mathbf{x};\hat{\mathbf{x}}_{\text{D},n}^{(k)},N_{\text{e}}^{(k)}\mathbf{I}_{R_{\text{D}}}\right)}\mathbf{x}}{\sum_{\mathbf{x}\in\mathcal{Q}^{R_{\text{D}}}}{\frac{\alpha}{|\mathcal{Q}^{R_{\text{D}}}|}\,\mathcal{CN}\left(\mathbf{x};\hat{\mathbf{x}}_{\text{D},n}^{(k)},N_{\text{e}}^{(k)}\mathbf{I}_{R_{\text{D}}}\right)} + (1-\alpha)\mathcal{CN}\left(\mathbf{0};\hat{\mathbf{x}}_{\text{D},n}^{(k)},N_{\text{e}}^{(k)}\mathbf{I}_{R_{\text{D}}}\right)}.
\label{PME_0}
\end{equation}
\hrulefill
\end{figure*}
\setcounter{equation}{\value{TempEqCnt}}

\begin{figure*}
	\centering
\includegraphics[width=0.99\textwidth]{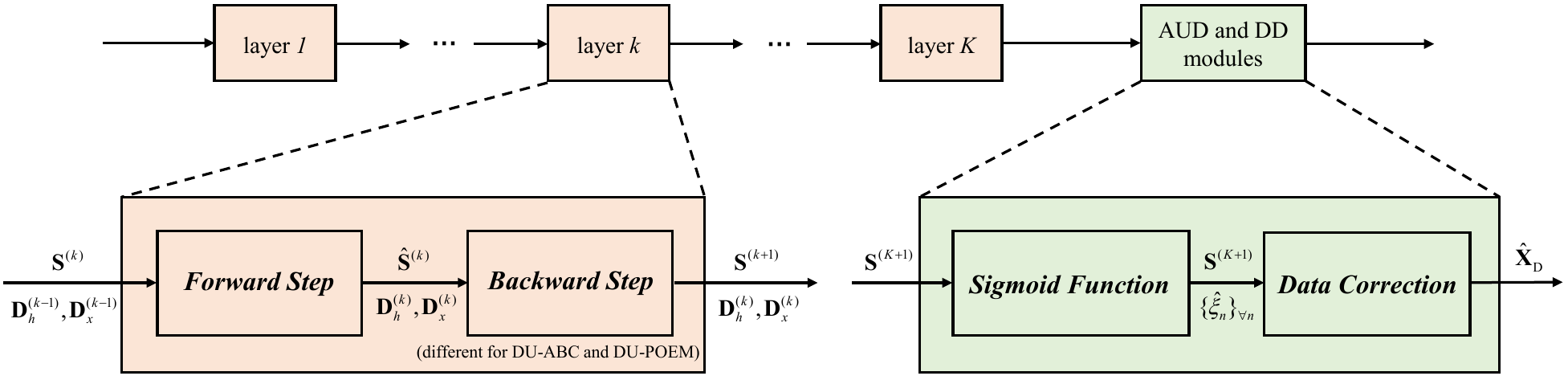}
	\caption{Architecture details of the DU-ABC and DU-POEM algorithms for JACD, differing only in the backward step.}
	\label{DU_architecture}
\end{figure*}

\subsection{PME-Based JACD Algorithm}\label{Section_IV_B}
An alternative technique that considers the discrete constellation constraint in $\mathbf{X}_{\text{D}}$ uses a PME, which denoises the data matrix $\mathbf{X}_{\text{D}}$; for this approach, we set $\lambda=0$. 
We now only discuss the differences to the box-constrained FBS algorithm from Section \ref{III_B} with $\lambda=0$, which are in the computation of the proximal operator for $\mathbf{X}_{\text{D}}$.

In our paper, the PME assumes that one observes a signal of interest through noisy Gaussian observations (see Definition~3 in Section \ref{Section_II_B}) and obtains the best estimates in terms of minimizing the mean square error by leveraging the known prior probability of the signal. 
As in~\cite{song2021soft,marti2023mitigating}, instead of calculating a proximal operator for the data matrix $\mathbf{X}_{\text{D}}$, we simply denoise the output of the forward step using a carefully designed PME. 
To this end, we model the $n$th UE's estimated data vector $\hat{\mathbf{x}}_{\text{D},n}^{(k)}$ in the $k$th iteration~as 
\begin{equation}
	\hat{\mathbf{x}}_{\text{D},n}^{(k)} = \mathbf{x}_{\text{D},n} + \mathbf{e}^{(k)}.
 \label{x_D_error}
\end{equation}
Here, $\mathbf{x}_{\text{D},n}$ is unknown and $\mathbf{e}^{(k)}\in\mathbb{R}^{R_{\text{D}}}$ is the estimation error at iteration $k$, which we assume  to be complex Gaussian following the distribution $\mathbf{e}^{(k)}\sim\mathcal{CN}(\mathbf{0}, N_{\text{e}}^{(k)}\mathbf{I}_{R_{\text{D}}})$. 
Here, the vector $\mathbf{x}_{\text{D},n}$ from \eqref{x_D_error} follows $\alpha$-mixed discrete uniform distribution on $\bar{\mathcal{Q}}$, i.e., $\mathbf{x}_{\text{D},n}\sim U_{\alpha,\bar{\mathcal{Q}}}\{\mathbf{x}_{\text{D},n}\}$ (see Definition~1 in Section \ref{Section_II_B}),  where $\alpha$ is the UE activity probability.
Accordingly, we can employ the PME of $\mathbf{x}_{\text{D},n}$ under the prior $U_{\alpha,\bar{\mathcal{Q}}}\{\mathbf{x}_{\text{D},n}\}$, i.e., $\text{PME}(\hat{\mathbf{x}}_{\text{D},n}^{(k)},\bar{\mathcal{Q}},U_{\alpha,\bar{\mathcal{Q}}}\{\mathbf{x}\},N_{\text{e}}^{(k)})$ (refer to Definition 3 in Section \ref{Section_II_B}) as the estimate of the data matrix $\mathbf{X}_{\text{D}}$~\cite{sun2023deep_unfolding}, expressed as in~\eqref{PME_0}. 
Algorithm~\ref{alg2} outlines the pseudocode for the proposed PME-based JACD~algorithm.

\begin{algorithm}[t]
	\SetAlgoNoLine
	\caption{PME-Based JACD Algorithm}
	\label{alg2}
	\textbf{Input}: $\mathbf{Y}$, $\mathbf{X}_{\text{P}}$, $\mu_h$, $\mu_x$, and $K$.\\
 \textbf{Initialization}: $\mathbf{S}^{(1)}=[(\mathbf{H}^{(1)})^H, \mathbf{X}_{\text{D}}^{(1)}]^H$.\\
 \For{$k=1,\ldots,K$}{
 \textit{Forward Step:} Calculate $\hat{\mathbf{S}}^{(k)}$ via \eqref{forward} and \eqref{gradient}.\\
 \textit{Backward Step:} Calculate $\mathbf{H}^{(k+1)}$ and $\mathbf{X}_{\text{D}}^{(k+1)}$ via \eqref{Shrinkage} and \eqref{PME_0}, respectively.
 }
	\textbf{Output}: $\mathbf{H}^{(K+1)}$ and $\mathbf{X}_{\text{D}}^{(K+1)}$.
\end{algorithm}

\subsection{Active User Detection and Data Detection}
After estimating the channel and data matrices over $ K $ iterations using box-constrained FBS or PME-based JACD, we proceed with AUD and DD for both algorithms. 
As in~\cite{sun2023joint_arxiv}, we determine UE activity based on the channel energy. 
Specifically, if the channel energy of the $n$th UE surpasses a threshold $T_{\text{AUD}}$, then the UE is deemed active; otherwise, we consider it inactive. 
Mathematically, we describe AUD as follows:
\setcounter{equation}{27}
\begin{equation}
    \hat{\xi}_n=\mathbb{I}\left\{\left\|\mathbf{h}_n^{(K+1)}\right\|_F^2\ge T_{\text{AUD}}\right\}.
    \label{AUD_threshold}
\end{equation}
The estimated UE activity indicators $\{\hat{\xi}_n\}_{\forall n}$ can now be used to update the estimated data matrix as
\begin{equation}
\hat{\mathbf{X}}_{\text{D}}=\text{diag}\left\{\hat{\xi}_1,\ldots,\hat{\xi}_N\right\}\tilde{\mathbf{X}}_{\text{D}},
\label{hat_X_D}
\end{equation}
where $\tilde{\mathbf{X}}_{\text{D}}\in\mathcal{Q}^{N\times R_{\text{D}}}$ is obtained by mapping the entries in~$\mathbf{X}_{\text{D}}^{(K+1)}$ to the nearest symbols in $\mathcal{Q}$ as follows: 
\begin{equation}
\tilde{\mathbf{X}}_{\text{D}}=\mathop{\arg\min}_{\scriptstyle{\mathbf{X}\in\mathcal{Q}^{N\times R_{\text{D}}}}}\left\|\mathbf{X}-\mathbf{X}_{\text{D}}^{(K+1)}\right\|_F^2.
\label{tilde_X_D}
\end{equation}

In Section \ref{IV-C}, we will introduce a trainable soft-output AUD module that leverages information from both the estimated channel and data matrices to improve  AUD performance.

\subsection{Complexity Comparison}
We assess the computational complexity of our algorithms using the number of complex-valued multiplications, which are expressed in big-O notation.
For each iteration, the complexity of the forward steps in these algorithms is $O(MPNR+MPNR_{\text{D}}+\mathbb{I}\{\lambda\ne0\}NR_{\text{D}}\mathcal{C}_1)$. 
Here, $\mathcal{C}_1$ varies depending on the chosen regularizer\footnote{For instance, $\mathcal{C}_1=1$ for the regularizer $\mathcal{C}\left(\mathbf{X}_{\text{D}}\right)=-\|\mathbf{X}_{\text{D}}\odot\mathbf{X}_{\text{D}}^*-B^2\mathbf{1}_{N\times R_{\text{D}}}\|_F^2$.}.
The per-iteration complexities of the backward steps in the box-constrained FBS and PME-based JACD algorithms are $O(MPN + N R_{\text{D}} )$ and $O( MPN + NR_{\text{D}}|\mathcal{Q}|^{R_{\text{D}}})$, respectively. 
In addition, the computational complexity of the AUD and DD module is $O(MPN+NR_{\text{D}})$.

The summation terms in the PME expression~\eqref{PME_0} lead to exponential complexity with the data length $R_{\text{D}}$ serving as the exponent, which results in higher complexity for the PME-based JACD algorithm compared to the box-constrained FBS algorithm.
In Section \ref{Section_V}, we will leverage DU to tune hyper-parameters in these algorithms, improving their effectiveness automatically. 
Besides that, we introduce approximations to replace the costly computations in \eqref{KKT_solution} and \eqref{PME_0} in the backward steps of these algorithms to further reduce~complexity.

\section{Algorithm Tuning Using Deep-Unfolding}\label{Section_V}
The JACD algorithms introduced in Section~\ref{Section_IV} involve numerous hyper-parameters, making manual parameter tuning challenging. We apply DU to tune all of the involved hyper-parameters automatically. 
The resulting deep-unfolded algorithms are called the Deep-Unfolding-based Approximate Box-Constrained (DU-ABC) algorithm and the Deep-Unfolding-based aPproximate pOsterior mEan estiMator (DU-POEM) algorithm. Their corresponding deep-unfolded architecture is outlined in Fig.~\ref{DU_architecture}.

\setcounter{TempEqCnt}{\value{equation}}
\setcounter{equation}{39}
\begin{figure*}
\vspace{-1ex}
\begin{equation}
 \mathbf{e}_r^T \text{PME}\left(\hat{\mathbf{x}}_{\text{D},n}^{(k)},\mathcal{Q}^{R_{\text{D}}}, U_{\mathcal{Q}^{R_{\text{D}}}}\{\mathbf{x}\},N_{\text{e}}^{(k)}\right) =\text{PME}\left(\hat{\mathbf{x}}_{\text{D},n}^{(k)}(r),\mathcal{Q}, U_{\mathcal{Q}}\{x\},N_{\text{e}}^{(k)}\right)=\frac{\sum_{x\in\mathcal{Q}}\mathcal{CN}\left(x;\hat{\mathbf{x}}_{\text{D},n}^{(k)}(r),N_{\text{e}}^{(k)}\right)x}{\sum_{x\in\mathcal{Q}}\mathcal{CN}\left(x;\hat{\mathbf{x}}_{\text{D},n}^{(k)}(r),N_{\text{e}}^{(k)}\right)},
\label{PME_single_element}
\end{equation}
\hrulefill
\vspace{-3ex}
\end{figure*}
\setcounter{equation}{\value{TempEqCnt}}

\subsection{DU-ABC Algorithm}
\subsubsection{Forward Step} 
In the forward step of the box-constrained FBS algorithm, the same step size $\tau^{(k)}$ in each iteration is utilized to compute both $\hat{\mathbf{H}}^{(k)}$ and $\hat{\mathbf{X}}_{\text{D}}^{(k)}$. 
Due to the vast difference in dynamic range of $\mathbf{H}$ and $\mathbf{X}_{\text{D}}$, for DU-ABC, we introduce separate step sizes $\tau_h^{(k)}$ and $\tau_x^{(k)}$ to update $\hat{\mathbf{H}}^{(k)}$ and $\hat{\mathbf{X}}_{\text{D}}^{(k)}$, respectively. 
To accelerate convergence, we apply a momentum strategy, where the gradient information of all previous iterations is used to compute~$\hat{\mathbf{S}}$ in each iteration~\cite{sun2023deep_unfolding,marti2023mitigating}. 
Furthermore, we allow the penalty coefficient $\lambda$ to be iteration-dependent as $\{\lambda^{(k)}\}_{\forall k}$. 
In summary, the forward step~\eqref{forward} of the box-constrained FBS algorithm is modified~as 
\begin{align}
		\hat{\mathbf{H}}^{(k)}&= \mathbf{H}^{(k)}+\mathbf{D}_h^{(k)},\label{Momentum1-1}\\
		 \hat{\mathbf{X}}_{\text{D}}^{(k)}&= \mathbf{X}_{\text{D}}^{(k)}+\mathbf{D}_x^{(k)},
\label{Momentum1-2}
\end{align}
where the momentum terms $\mathbf{D}_h^{(k)}$ and~$\mathbf{D}_x^{(k)}$ incorporate the gradient information from the first $ k $ iterations, and are given~by 
\begin{align}
	\!\!\!\mathbf{D}_h^{(k)}  &=   \tau_h^{(k)}  \left(\mathbf{Y} - \mathbf{H}^{(k)}\mathbf{X}^{(k)}\right)\left(\mathbf{X}^{(k)}\right)\!^H  +  \eta_h^{(k)} \mathbf{D}_h^{(k-1)},\label{Momentum2-1}\\
	\!\!\!\mathbf{D}_x^{(k)}  &=   \tau_x^{(k)} \left( \left(\mathbf{Y}_{\text{D}}-\mathbf{H}^{(k)}\mathbf{X}_{\text{D}}^{(k)}\right)\! ^H\mathbf{H}^{(k)} + \lambda^{(k)} \frac{\partial \mathcal{C} \left(\mathbf{X}_{\text{D}}\right) }{\partial\mathbf{X}_{\text{D}}^T}\right)\nonumber\\
 &\quad\, + \eta_x^{(k)} \mathbf{D}_x^{(k-1)}.\label{Momentum2-2}
\end{align}
Here, $\eta_h^{(k)}$ and $\eta_x^{(k)}$ are weights for the momentum terms of the quantities $\hat{\mathbf{H}}^{(k)}$ and~$\hat{\mathbf{X}}_{\text{D}}^{(k)}$, respectively, and $\mathbf{D}_h^{(0)}=\mathbf{0}$
and $\mathbf{D}_x^{(0)}=\mathbf{0}$. 

\subsubsection{Backward Step}
{Since we introduce separate step sizes $\tau_h^{(k)}$ and $\tau_x^{(k)}$ in the forward step of DU-ABC, we accordingly define trainable parameters $\tilde{\mu}_h^{(k)} \triangleq \tau_h^{(k)}\mu_h$ and $\tilde{\mu}_x^{(k)} \triangleq \tau_x^{(k)}\mu_h$ for all $k$, to facilitate subsequent computations. 
Consequently, the proximal operator for $\mathbf{H}$ in \eqref{Shrinkage} is modified as follows:
\begin{equation}
 \mathbf{h}_{n,p}^{(k+1)} = \textsf{Shrinkage}\left(\hat{\mathbf{h}}_{n,p}^{(k)},\tilde{\mu}_h^{(k)}\right).
	\label{Shrinkage_untie}
\end{equation}
As for the proximal operation on $\mathbf{X}_{\text{D}}$, the optimal solution of the problem~\eqref{xd_prox}  involves a complicated quartic equation, as mentioned in Proposition~1, thereby preventing the utilization of DU techniques. 
To solve this problem, we introduce a simpler alternative: first solve problem \eqref{xd_prox} without considering the constraints to obtain the optimal solution $\textsf{Shrinkage}\big(\hat{\mathbf{x}}_{\text{D},n}^{(k)},\tilde{\mu}_x^{(k)}\big)$; then, clamp the result to the convex hull $\mathcal{B}$. The procedure is 
\begin{equation}
	\textstyle \mathbf{x}_{\text{D},n}^{(k+1)} \!=\! \textsf{Clamp} \big(\omega^{(k)}\textsf{Shrinkage}\big(\hat{\mathbf{x}}_{\text{D},n}^{(k)},\tilde{\mu}_x^{(k)}\big) \!+\! \mathbf{b}^{(k)},-B,B\big) ,
\end{equation}
with the clamp function defined in \eqref{clamp_definition}.
In addition, $\omega^{(k)}\in\mathbb{R}$ and $\mathbf{b}^{(k)}\in\mathbb{C}^{R_{\text{D}}}$ are introduced trainable parameters for the coefficient and bias vector at the $k$th iteration, respectively, to increase the flexibility of optimization.

For DU-ABC, the trainable hyper-parameters in forward and backward steps are $\tau_h^{(k)},\eta_h^{(k)},\tau_x^{(k)},\eta_x^{(k)},\lambda^{(k)},\Theta_{\mathcal{C}}^{(k)}$ and $\tilde{\mu}_h^{(k)},\tilde{\mu}_x^{(k)},\omega ^{(k)},\mathbf{b}^{(k)}$, respectively, where $\Theta_{\mathcal{C}}^{(k)}$ denotes the set of trainable hyper-parameters in the regularizer $\mathcal{C}(\mathbf{X}_{\text{D}})$.

\subsection{DU-POEM Algorithm}\label{Section_V_B}
The forward step and the proximal operations on $\mathbf{H}$ in the backward step of the DU-POEM algorithm align with those of the DU-ABC algorithm with $\lambda^{(k)}=0,\forall k$, as detailed in \eqref{Momentum1-1}-\eqref{Shrinkage_untie}. 
We now only focus on the proximal operation for $\mathbf{X}_{\text{D}}$ in DU-POEM, which is different from that of the DU-ABC.

As for the PME of $\mathbf{X}_{\text{D}}$ in the PME-based JACD algorithm, the summation of numerous exponential terms in \eqref{PME_0} can result in high computational complexity and lead to numerical stability issues. 
To mitigate these issues, we show the following proposition that reveals the linear relationship between the PME of $\mathbf{x}_{\text{D},n}$ under two specific prior distributions: $U_{\alpha,\bar{\mathcal{Q}}}\{\mathbf{x}_{\text{D},n}\}$ and $U_{\mathcal{Q}^{R_{\text{D}}}}\{\mathbf{x}_{\text{D},n}\}$ (refer to Definition 2 in Section \ref{Section_II_B}).

\textit{\textbf{Proposition 2:} 
Given observation vector $\hat{\mathbf{x}}=\mathbf{x} +\mathbf{e}$ with Gaussian estimation error $\mathbf{e}\sim\mathcal{CN}\left(\mathbf{0},N_{\text{e}}\mathbf{I}\right)$, there is a linear relationship between $\text{PME}(\hat{\mathbf{x}},\bar{\mathcal{S}},U_{\alpha,\bar{\mathcal{S}}}\{\mathbf{x}\},N_{\text{e}})$ and $\text{PME}(\hat{\mathbf{x}},\mathcal{S},U_{\mathcal{S}}\{\mathbf{x}\},N_{\text{e}})$,~i.e., 
\begin{equation}
\begin{aligned}
    &\text{PME}(\hat{\mathbf{x}},\bar{\mathcal{S}},U_{\alpha,\bar{\mathcal{S}}}\{\mathbf{x}\},N_{\text{e}})\\
    &=C_{\text{PME}}(\hat{\mathbf{x}},\mathcal{S},\alpha,N_{\text{e}})\,\text{PME}(\hat{\mathbf{x}},\mathcal{S},U_{\mathcal{S}}\{\mathbf{x}\},N_{\text{e}}),
\end{aligned}
    \label{linear_PME}
\end{equation}
where $\bar{\mathcal{S}}=\{\mathcal{S},\mathbf{0}\}$, and the coefficient $C_{\text{PME}}(\hat{\mathbf{x}},\mathcal{S},\alpha,N_{\text{e}})$ is defined as
\begin{equation}
\!\!\textstyle C_{\text{PME}}(\hat{\mathbf{x}},\mathcal{S},\alpha,N_{\text{e}})\!=\!\alpha\left(\alpha\!+\!\left(1\!-\!\alpha\right)\!\frac{{|\mathcal{S}|}\,\mathcal{CN}(\mathbf{0};\hat{\mathbf{x}},N_{\text{e}}\mathbf{I})}{\sum_{\mathbf{x}\in\mathcal{S}}\mathcal{CN}(\mathbf{x};\hat{\mathbf{x}},N_{\text{e}}\mathbf{I})}\right)^{-1}\!\!\!.
\label{C_PME}
\end{equation}}
The proof is given in Appendix \ref{proof of Proposition 2}.

According to Proposition 2, we can reformulate the PME of $\mathbf{x}_{\text{D},n}$ under the prior $U_{\alpha,\bar{\mathcal{Q}}}\{\mathbf{x}\}$, $\text{PME}(\hat{\mathbf{x}}_{\text{D},n}^{(k)},\bar{\mathcal{Q}},U_{\alpha,\bar{\mathcal{Q}}}\{\mathbf{x}\},N_{\text{e}}^{(k)})$, in equation \eqref{PME_0} as the product of a coefficient and $\text{PME}(\hat{\mathbf{x}}_{\text{D},n}^{(k)},\mathcal{Q}^{R_{\text{D}}}, U_{\mathcal{Q}^{R_{\text{D}}}}\{\mathbf{x}\},N_{\text{e}}^{(k)})$.
The main advantage of this reformulation is that we can further decouple each element in $\text{PME}(\hat{\mathbf{x}}_{\text{D},n}^{(k)},\mathcal{Q}^{R_{\text{D}}}, U_{\mathcal{Q}^{R_{\text{D}}}}\{\mathbf{x}\},N_{\text{e}}^{(k)})$ and compute them independently, which is explained by Proposition 3. 
The proof is given in Appendix \ref{proof of Proposition 3}.

\textit{\textbf{Proposition 3:} 
Given observation vector $\hat{\mathbf{x}}=\mathbf{x} +\mathbf{e}\in\mathbb{C}^S$ with $\mathbf{x}\sim U_{\mathcal{S}}\{\mathbf{x}\}$ and Gaussian estimation error $\mathbf{e}\sim\mathcal{CN}\left(\mathbf{0},N_{\text{e}}\mathbf{I}_S\right)$, we can express the $s$-th entry of $\text{PME}(\hat{\mathbf{x}},\mathcal{S}, U_{\mathcal{S}}\{\mathbf{x}\},N_{\text{e}})$~as
\begin{equation}
\!\!\!\mathbf{e}_s^T\text{PME}(\hat{\mathbf{x}},\mathcal{S}, U_{\mathcal{S}}\{\mathbf{x}\},N_{\text{e}})
    \!=\!\text{PME}(\hat{\mathbf{x}}(s),\mathcal{R}, U_{\mathcal{R}}\{x\},N_{\text{e}}),\!
\end{equation}
where $\mathcal{S}=\mathcal{R}^S$.
}

According to Proposition 3, we can calculate the $r$th entry of $\text{PME}(\hat{\mathbf{x}}_{\text{D},n}^{(k)},\mathcal{Q}^{R_{\text{D}}}, U_{\mathcal{Q}^{R_{\text{D}}}}\{\mathbf{x}\},N_{\text{e}}^{(k)})$ by the expression \eqref{PME_single_element}, 
which only relates to $\hat{\mathbf{x}}_{\text{D},n}^{(k)}(r)$ and avoids a summation of a large number of exponential terms, thereby reducing complexity and avoiding numerical stability issues.

\begin{figure}[tp]
	\centering
\includegraphics[width=0.48\textwidth]{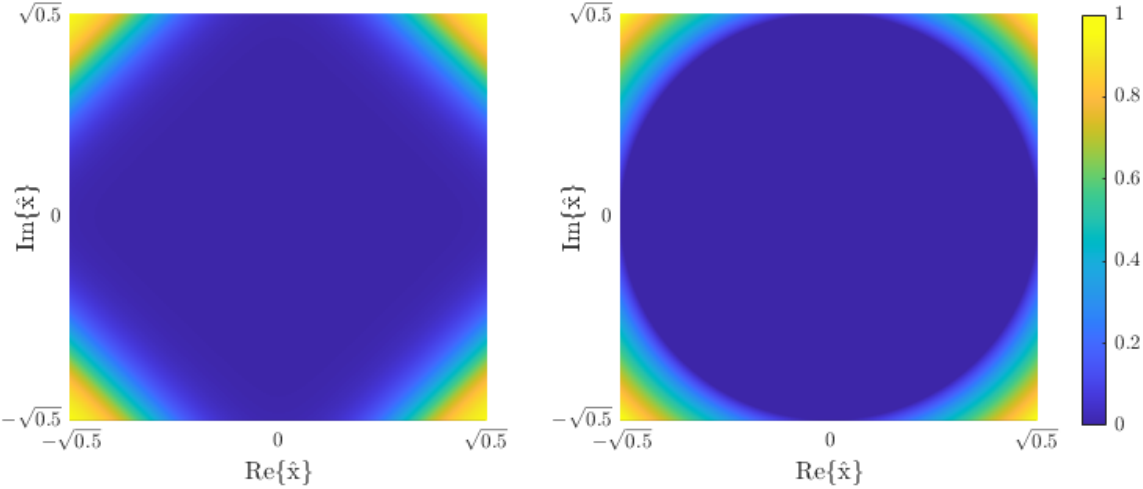}
	\caption{Diagrams of the coefficient $C_{\text{PME}}(\hat{x},\mathcal{S},\alpha,N_{\text{e}})$ with $\mathcal{S}=\{\pm \sqrt{0.5} \pm j\sqrt{0.5}\}$, $\alpha=0.02$, and $N_e=0.12$ (on the left) and its approximation $C_{\text{APME}}(\hat{x},\rho,\nu)$ with $\rho=3.49$ and $\nu=2.46$ (on the right).}
	\label{C_APME}
\end{figure}

Although $\text{PME}(\hat{\mathbf{x}}, \mathcal{S}, U_{\mathcal{S}}\{\mathbf{x}\}, N_{\text{e}})$ in \eqref{linear_PME} requires low computational complexity and alleviates numerical stability issues, Proposition~2 also introduces $C_{\text{PME}}(\hat{\mathbf{x}}, \mathcal{S}, \alpha, N_{\text{e}})$ as shown in \eqref{C_PME}, which remains complex due to the summation of numerous exponential terms in the denominator. 
To address this, we propose the following approximate shrinkage operation as a simplified alternative:
\setcounter{equation}{40}
\begin{equation}
\label{approximate_shrinkage_definition}
	\begin{aligned}
	  \!\!\!C_{\text{APME}}(\hat{\mathbf{x}},\rho,\nu)\!=\! \left\{\!\!\begin{array}{l}
			\textsf{Clamp}\Big(\rho\!-\!\frac{\nu}{\|\hat{\mathbf{x}}\|_F},0,1\Big),\! \text{ if }\hat{\mathbf{x}}\neq \mathbf{0},\\
		0,\! \text{ if }\hat{\mathbf{x}}=\mathbf{0}.
		\end{array}\right.
	\end{aligned}
\end{equation}
where $\rho$ and $\nu$ are tunable parameters.
In Fig.~\ref{C_APME}, we illustrate $C_{\text{PME}}(\hat{x},\mathcal{S},\alpha, N_{\text{e}})$ with $\mathcal{S}=\{\pm \sqrt{0.5} \pm j\sqrt{0.5}\}$, $\alpha=0.02$, and $N_e=0.12$ alongside its approximation $C_{\text{APME}}(\hat{x},\rho,\nu)$ with $\rho=3.49$ and $\nu=2.46$ in a simplified one-dimensional complex-value space $\hat{x}\in\mathbb{C}$ for ease of visualization.
Evidently, the resulting approximation  $C_{\text{APME}}(\hat{x},\rho,\nu)$, illustrated on the right of Fig.~\ref{C_APME}, exhibits sufficient similarity with $C_{\text{PME}}(\hat{x},\mathcal{S},\alpha, N_{\text{e}})$.

Consequently, we employ an approximate PME at the backward step to replace the exact PME \eqref{PME_0} as done in~\cite{sun2023deep_unfolding}:
\begin{equation}
\begin{aligned}
\mathbf{x}_{\text{D},n}^{(k+1)}&=C_{\text{APME}}\left(\hat{\mathbf{x}}_{\text{D},n}^{(k)},\rho^{(k)},\nu^{(k)}\right)\\
&\quad\times \text{PME}\left(\hat{\mathbf{x}}_{\text{D},n}^{(k)},\mathcal{Q}^{R_{\text{D}}}, U_{\mathcal{Q}^{R_{\text{D}}}}\{\mathbf{x}\},N_{\text{e}}^{(k)}\right).
\end{aligned}
\end{equation}
Here, the hyper-parameters $\rho^{(k)},\nu^{(k)}$ and the variance of the estimation error $N_{\text{e}}^{(k)}$ at iteration $k$ are trainable.

In DU-POEM, the trainable hyper-parameters in the forward and backward steps are $\tau_h^{(k)},\eta_h^{(k)},\tau_x^{(k)},\eta_x^{(k)}$ and $\tilde{\mu}_h^{(k)}, N_{\rm{e}}^{(k)},\rho^{(k)}, \nu^{(k)}$, respectively.

\subsection{Trainable Soft-Output Active User Detection and Data Detection Modules}\label{IV-C}
Both the sparsity in the channel matrix and in the data matrix indicate UE activity.
To obtain accurate soft information for UE activity $\{L_n\}_{\forall n}$, we propose the use of a sigmoid function as in~\cite{sun2023deep_unfolding} to fully extract activity information from both the channel and data matrices as follows:
\begin{equation}
	L_n \!=\! \Big(1\!+\exp\!\Big(T_{\text{th}}\!-\omega_h\left\|\mathbf{h}_n^{(K+1)}\right\|_F^2\!\!-\omega_x\left\|\mathbf{x}_{\text{D},n}^{(K+1)}\right\|_F^2\!\Big)\!\Big)^{-1}\!\!\!.
\end{equation}
Here, the parameters $\omega_h$, $\omega_x$, and $T_{\text{th}}$ are tuned using DU. 
Utilizing soft information, we can detect the UEs' active states by comparing them against a threshold $\bar{L}\in[0,1]$, i.e., 
\begin{equation}
    \hat{\xi}_n = \mathbb{I}\left\{L_n>\bar{L}\right\}.
\end{equation}
Determining $\bar{L}$ depends on the desired UE miss-detection and false-detection rates\footnote{The UE miss-detection rate is the ratio of the number of active UEs mistakenly deemed inactive to the total number of UEs. 
The UE false-detection rate is the ratio of the number of inactive UEs incorrectly classified as active to the total number of UEs.}. 
Generally, larger $\bar{L}$ might result in more UEs being detected as inactive, subsequently potentially increasing the UE miss-detection rate and reducing the UE false-detection rate. 
Strictly speaking, as $\bar{L}$ increases, the UE miss-detection rate should not decrease, and the UE false-detection rate should not increase.

The DD modules of the DU-ABC and DU-POEM algorithms are the same as \eqref{hat_X_D} and \eqref{tilde_X_D}.

\subsection{Training Procedure}
In our previous work~\cite{sun2023deep_unfolding}, we trained the hyper-parameters in the unfolded layers of these algorithms and the AUD module separately.
As a result, the AUD module is unable to guide these unfolded layers to estimate more accurate outputs that could potentially improve AUD performance. 
This one-way interaction limits the effectiveness of the parameter tuning.
Furthermore, the performance of these unfolded layers was also not fully optimized in~\cite{sun2023deep_unfolding} due to the lack of valuable feedback from the AUD module. 
These separate training processes result in a missed opportunity for more effective hyper-parameter tuning.
To achieve better JACD performance, we choose to jointly train the hyper-parameters in the unfolded layers in conjunction with the AUD module utilizing the following loss~function:
\begin{equation}
\operatorname{Loss}=\left\|\operatorname{diag}\left\{L_1, \ldots, L_N\right\}\mathbf{X}_{\text{D}}^{(K+1)}-\mathbf{X}_{\text{D}}\right\|_F^2.
\label{loss}
\end{equation}
The above loss function underscores our emphasis on the precision of both AUD and DD performance, aligning seamlessly with the objectives of the practical wireless communication systems. 
We note that due to the absence of the error term comparing the estimated channel matrix and actual channel matrix in the loss function \eqref{loss}, i.e., we do not explicitly optimize our algorithms for CE accuracy.

Note that the shrinkage operation \eqref{shrinkage_definition} and approximate shrinkage operation \eqref{approximate_shrinkage_definition} applied in DU-based algorithms have no gradient at $\mathbf{0}$. 
To circumvent this issue, we replace the denominator $ \|\mathbf{x}\|_F $ in these equations with $ \sqrt{\|\mathbf{x}\|_F^2+\epsilon} $, producing valid gradients and avoiding a denominator~of~$0$, where $\epsilon$ is a small value (we use $\epsilon=10^{-40}$ in the simulations).

\subsection{Complexity Comparison}
The computational complexity per iteration for the forward steps in both the DU-ABC and DU-POEM algorithms is $O(MPNR+MPNR_{\text{D}}+\mathbb{I}\{\lambda\ne0\}NR_{\text{D}}\mathcal{C}_1)$. 
Additionally, for the backward steps, the per-iteration complexity is $O(MPN + N R_{\text{D}} )$ for the DU-ABC algorithm and $O( MPN+NR_{\text{D}}|\mathcal{Q}|)$ for the DU-POEM algorithm. The computational complexity of the trainable AUD and DD module is $O(MPN+NR_{\text{D}})$.

By approximating the PME expression~\eqref{PME_0} through Propositions 2 and 3, we have significantly reduced the complexity of the backward step in the DU-POEM algorithm to $O( MPN+NR_{\text{D}}|\mathcal{Q}|)$, in contrast to $O( MPN+NR_{\text{D}}|\mathcal{Q}|^{R_{\text{D}}})$ in the PME-based JACD algorithm. 
This reduction makes the computational complexities of both the DU-ABC and DU-POEM algorithms comparable, improving their efficiency and~practicality.

\section{Simulation Results}\label{Section_VI}

We now demonstrate the efficacy of our proposed JACD algorithms and compare them to existing baseline methods.

\subsection{Simulation Setup}
Building upon the system settings from~\cite{sun2023joint_arxiv,sun2023deep_unfolding,song2022joint}, we consider a cell-free wireless communication system in an area of 500\,m $\times$ 500\,m. 
Unless stated otherwise, we use the following assumptions.
We consider $P = 60$ uniformly distributed APs at the height of $h_{\text{AP}} = 15$\,m, each with $M = 4$ antennas, that serve $N = 400$ uniformly distributed single-antenna UEs at the height of $h_{\text{UE}}=1.65$\,m.
We set the UE activity probability to $\alpha = 0.2$.
Active UEs transmit $R_{\text{P}} = 50$ pilot signals, originating from a complex equiangular tight frame as described in~\cite{Tropp2005designing}, and $R_{\text{D}} = 200$ QPSK data signals over the channel with a bandwidth of $20$\,MHz and a carrier frequency of $f_c = 1900$\,MHz.
These signals satisfy the energy constraints $\|\mathbf{x}_{\text{P},n}\|_F^2=R_{\text{P}}$ and $\mathbb{E}\{\|\mathbf{X}_{\text{D}}(n,r)\|_F^2\}=1$,  implying that $B=\sqrt{0.5}$ for QPSK.
We assume that the UEs' transmission power is 0.1\,W, with power control allowing for a dynamic range of up to 12\,dB between the strongest and weakest UE~\cite{song2022joint}. 
Moreover, we account for a shadow fading variance of $\sigma_{\text{sh}}^2 = 8$\,dB, a noise figure of $9$\,dB, and a noise temperature of $290$\,K.

Regarding the channel model, we assume that the small-scale fading parameters follow the standard complex Gaussian distribution, while the large-scale fading follows a three-slope path-loss model~\cite{song2022joint,ngo2017cell,tang2001mobile}. 
Specifically, the large-scale fading between the $n$th UE and the $p$th AP, denoted as $\beta_{n,p}$, is given by~\cite[Eq. (52)]{ngo2017cell},~\cite{song2022joint},~\cite{tang2001mobile}
\begin{equation}
\!\!\beta_{n,p} \!=\! \left\{\!\!{\begin{array}{l}
				10^{\frac{-L+z_{n,p}}{10}}(d_{n,p})^{-3.5},\;\text{if}\; d_{n,p} > D_1,\\
				10^{-\frac{L}{10}}(d_{1})^{-1.5}(d_{n,p})^{-2},\;\text{if}\; D_0 < d_{n,p} \le D_1,\\
				10^{-\frac{L}{10}}(d_{1})^{-1.5}(d_{0})^{-2},\;\text{if}\; d_{n,p} \le D_0.
		\end{array}} \right.
		\label{pathloss}
	\end{equation}
Here, $d_{n,p}$\,[km] is the distance between the $n$th UE and the $p$th AP with breakpoints at $D_0 = 0.01$\,km and $D_1 = 0.05$\,km~\cite{ngo2017cell}. 
Besides, the shadow fading follows $z_{n,p}\sim\mathcal{N}(0,\sigma_{\text{sh}}^2)$, and $L\triangleq 45.5 + 35.46\log_{10}(f_c) -13.82\log_{10}(h_{\text{AP}})-(1.1\log_{10}(f_c)-0.7)h_{\text{UE}}$~\cite[Eq. (53)]{ngo2017cell}.

\subsection{Baseline Methods}
To assess the effectiveness of our algorithms\footnote{Since the PME-based JACD algorithm involves tuning numerous hyper-parameters and has~a~high computational complexity, this section only presents performance simulations for its deep-unfolded variant, DU-POEM, alongside the box-constrained FBS algorithm and DU-ABC.}, we introduce the following baseline methods for comparison: 
\begin{itemize}
\item \textbf{Baseline 1:} In this baseline, we first employ the FBS method \cite{goldstein2014field} to estimate sparse channels from the system model $\mathbf{Y}_{\text{P}} = \mathbf{H} \mathbf{X}_{\text{P}} + \mathbf{N}$. 
Then, active UEs are identified by equation \eqref{AUD_threshold} based on the estimated channels $\tilde{\mathbf{H}}_{\text{tmp}}$, resulting in the active UE set $\{\hat{\xi}_1, \ldots, \hat{\xi}_N\}$. 
Subsequently, we perform DD through $\tilde{\mathbf{X}}_{\text{D}} = (\tilde{\mathbf{H}}^H \tilde{\mathbf{H}})^{-1} \tilde{\mathbf{H}}^H \mathbf{Y}_{\text{D}}$, where $\tilde{\mathbf{H}} = \tilde{\mathbf{H}}_{\text{tmp}} \, \text{diag}\{\hat{\xi}_1, \ldots, \hat{\xi}_N\}$. 
Finally, we map the result $\tilde{\mathbf{X}}_{\text{D}}$ to the nearest constellation symbols using equation~\eqref{tilde_X_D}.

\item \textbf{Baseline 2:} In this baseline, we utilize the AMP algorithm \cite{chen2018sparse} to estimate sparse channels, while all other components remain unchanged from Baseline 1.

\item \textbf{Baseline 3:} This baseline retains all components of Baseline 1, except that we employ a soft MMSE-based iterative detection method \cite{zhang2019evaluation} for DD.

\item \textbf{Baseline 4:} We adopt a joint AUD-CE-DD method as proposed in \cite{zou2020alow}, which combines Bi-GAMP for CE and DD, alongside sum-product loopy belief propagation (LBP) for AUD.

\item \textbf{Baseline 5:} This baseline implements the FBS-based approach from \cite{song2022joint} for joint AUD-CE-DD, with active UEs identified using equation \eqref{AUD_threshold} based on the estimated~channels.
\end{itemize}

To accelerate convergence, we take the result of Baseline~1 to initialize Baseline~5, the box-constrained FBS, DU-ABC, and DU-POEM algorithms.
In addition, we carry out a maximum number of  $K=200$ iterations for {Baselines 1-to-5} and box-constrained FBS algorithm, with a stopping tolerance of $10^{-3}$ for FBS. 
We use $ K = 10$ layers (equal to the maximum number of iterations) for DU-ABC and DU-POEM.
To ensure a fair comparison with DU-ABC and DU-POEM, we also present the results of high-performance {Baselines 2, 4, 5} and the box-constrained FBS algorithm with only $10$ iterations.

To illustrate the trade-off between JACD performance and computational complexity, we now provide the computational complexities for Baselines 1-5 before we analyze their performance. 
Specifically, their computational complexities are $ O(MPNR_{\text{P}}K_{\text{iter}} + MPN^2 + N^3+ MPNR_{\text{D}}) $, $ O(MPNR_{\text{P}}K_{\text{iter}} + MPN^2 + N^3+ MPNR_{\text{D}}) $, $ O(MPNR_{\text{P}}K_{\text{iter}} + N^3R_{\text{D}}+N^2MPR_{\text{D}}) $, $ O(N^3K_{\text{iter}})$, and $ O(MPNRK_{\text{iter}}+MPNR_{\text{D}}K_{\text{iter}}) $, respectively, where $K_{\text{iter}}$ is the iteration number of these~baselines.

\subsection{Performance Metrics}
To evaluate the performance of the proposed algorithms and the baseline methods, we consider the following performance metrics: UE detection error rate (UDER), channel estimation normalized mean square error (NMSE), and average symbol error rate (ASER), which are defined as follows: 
\begin{align}
        &\textit{UDER}= \frac{\sum_{n=1}^N\left|\xi_n-\hat{\xi}_n\right|}{N},\\
        &\textit{NMSE}= \frac{\left\|\mathbf{H}-{\mathbf{H}}^{(K+1)}\right\|_F^2}{\left\|\mathbf{H}\right\|_F^2},\\        &\textit{ASER}=\frac{\sum_{n,r}\xi_n\mathbb{I}\left\{\mathbf{X}_\text{D}(n,r)\ne \tilde{\mathbf{X}}_\text{D}(n,r)\right\}}{R_{\text{D}}N_a}.
\end{align}
Furthermore, we also consider a receiver operating characteristic (ROC) curve analysis with the goal of exploring the trade-off between true positive rate (TPR) and false positive rate (FPR) for AUD, which are defined as follows:
\begin{align}
        \textit{FPR}&=\frac{\sum_{n=1}^N \mathbb{I}\{\xi_n=1,\hat{\xi}_n=1\}}{N_a},\\
        \textit{TPR}&= \frac{\sum_{n=1}^N \mathbb{I}\{\xi_n=0,\hat{\xi}_n=1\}}{N-N_a}.
\end{align}
The results shown next are from $5000$ Monte--Carlo trials.

\begin{figure*}\centering
	\subfigure[UDER]{\label{diff_P_UDER}\includegraphics[width=0.49\linewidth]{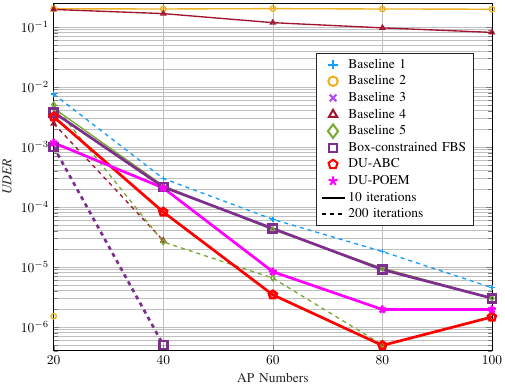}}
	\subfigure[NMSE]{\label{diff_P_NMSE}\includegraphics[width=0.49\linewidth]{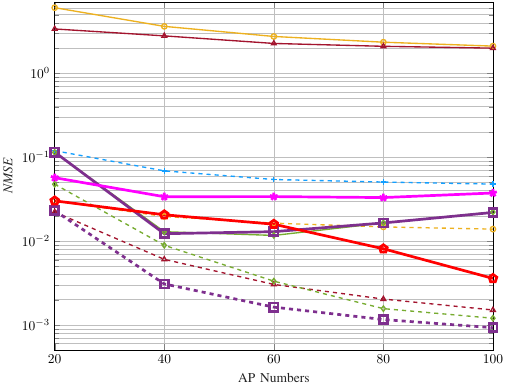}}
	\subfigure[ASER]{\label{diff_P_ASER}\includegraphics[width=0.49\linewidth]{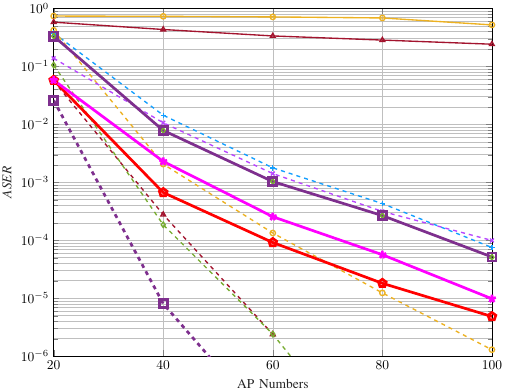}}
 \subfigure[ROC]{\label{ROC}\includegraphics[width=0.47\linewidth]{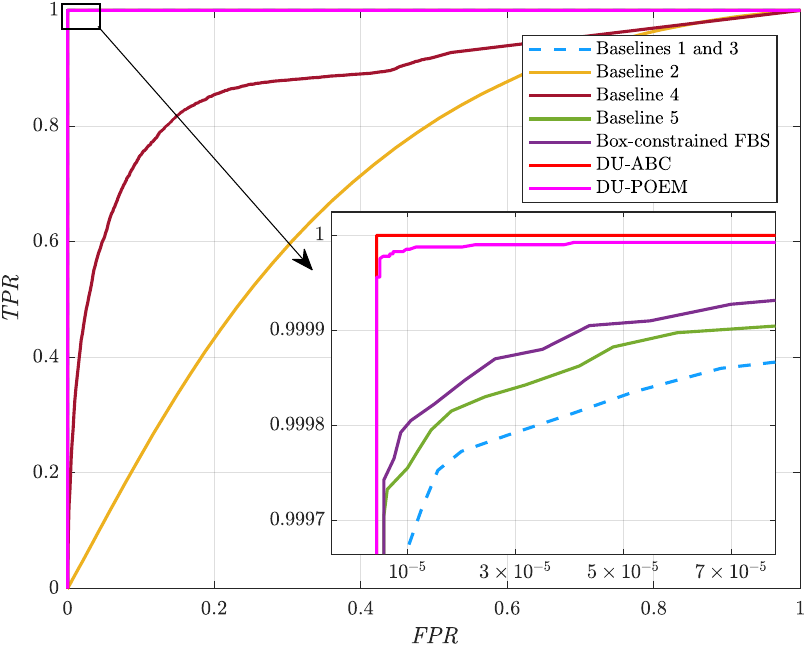}
}
	\caption{AUD, CE, and DD performance comparison for the number of APs $P=20,40,60,80,100$, as well as the ROC at $P=60$.}
\label{diff_P}
\end{figure*}

\subsection{Simulation Results}
\subsubsection{Performance Analysis for Different AP Numbers} 
In Fig.~\ref{diff_P}, we evaluate the JACD performance across various algorithms for different numbers of APs. 
Figs.~\ref{diff_P_UDER}-\ref{diff_P_ASER} show that, when running $200$ iterations, the box-constrained FBS algorithm consistently outperforms all other considered methods in terms of NMSE and ASER across various AP numbers. 
In addition, the performance of the box-constrained FBS algorithm in UDER surpasses that of most baseline methods. 
The superior performance of the box-constrained FBS algorithm is primarily due to its effective utilization of the block sparsity in the channel matrix and the row sparsity in the data matrix.

Furthermore, when executing $10$ iterations, DU-ABC and DU-POEM surpass all baseline methods in terms of ASER and UDER while maintaining comparable performance with others in NMSE.
The superiority of these DU-based algorithms in UDER and ASER can be attributed to the precise tuning of algorithm hyper-parameters through DU. 
However, their moderate performance in NMSE is due to the fact that their loss function primarily focuses on AUD and DD without accounting for CE accuracy.

\begin{figure*}
	\centering
	\subfigure[UDER]{\label{diff_alpha_UDER}\includegraphics[width=0.49\linewidth]{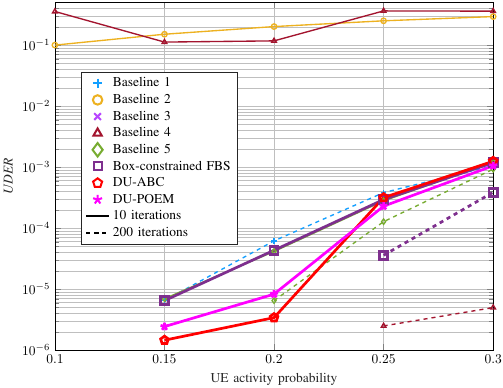}}
	\subfigure[NMSE]{\label{diff_alpha_NMSE}\includegraphics[width=0.49\linewidth]{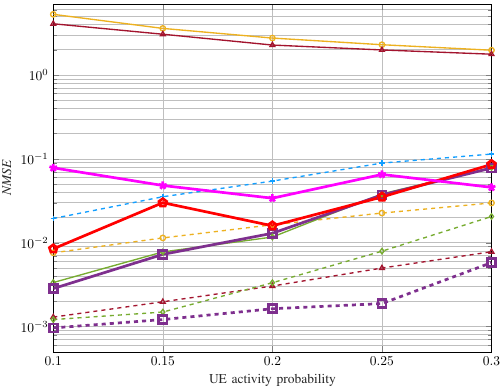}}
	\subfigure[ASER]{\label{diff_alpha_ASER}\includegraphics[width=0.49\linewidth]{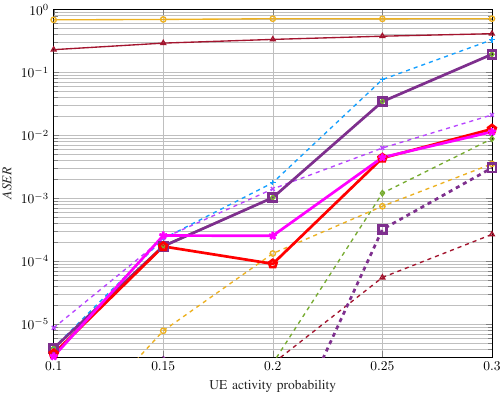}}
  \subfigure[ROC]{\label{ROC015}\includegraphics[width=0.47\linewidth]{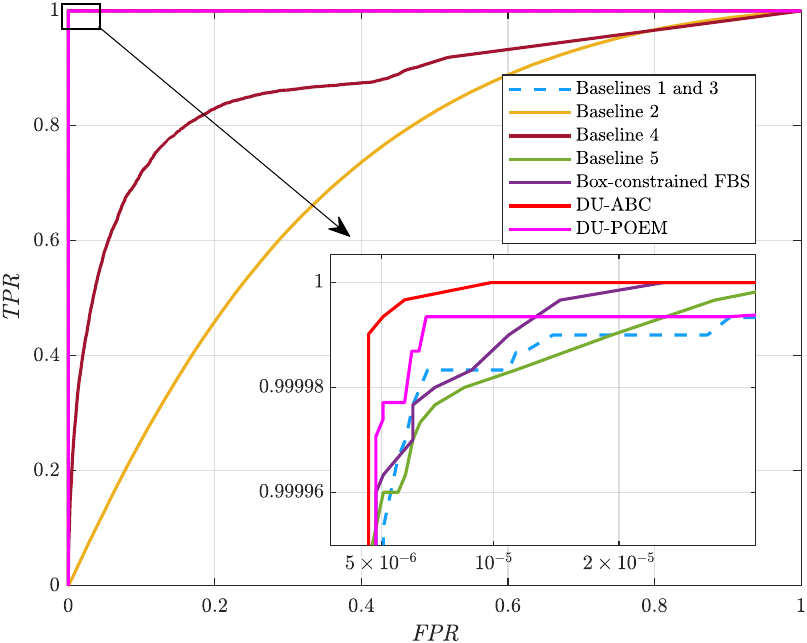}}
	\caption{AUD, CE, and DD performance comparison under activity probabilities $\alpha=0.1,0.15,0.2,0.25,0.3$, as well as the ROC at $\alpha=0.15$.}
	\label{diff_alpha}
\end{figure*}

The insufficient iterations for some algorithms to converge result in the following phenomena: (i) The NMSE of Baseline~5 and the box-constrained FBS algorithm with 10 iterations show instability as the number of APs increases, and (ii) AMP-based Baseline 2 and Bi-GAMP-based Baseline 4 experience significant JACD performance degradation when limited to 10 iterations compared to their performance with 200 iterations.
In addition, the UDER of DU-ABC at $ P=100 $ exhibits inferior performance compared to that at $ P=80 $, resulting from the optimization objective \eqref{loss} integrating both AUD and DD performance. 
In these scenarios, the optimization process might give precedence to the precision of DD and compromise the performance of AUD for a smaller value of the loss function.

In Fig.~\ref{ROC}, the ROC of various algorithms is depicted at $P=60$, where Baseline 1 and 3 run $200$ iterations while the remaining algorithms use only $10$ iterations. 
Notably, the AMP-based Baseline 2 and Bi-GAMP-based Baseline 4 perform the worst for $10$ iterations. 
In contrast, the DU-ABC and DU-POEM algorithms manifest superior AUD performance, as evidenced by their elevated TPR at the same FPR. 
For ROC, an increase in TPR often corresponds to a higher (or identical) FPR, which highlights that the selection of thresholds $\bar{L}$ and $T_{\text{AUD}}$ embodies a balance between TPR and FPR, contingent upon the targeted metrics for each. 

In summary, when running $200$ iterations, the proposed box-constrained FBS algorithm generally achieves the best AUD, CE, and DD performance across various numbers of APs. 
When limited to only $10$ iterations, the proposed DU-ABC and DU-POEM algorithms respectively demonstrate the best and second-best performance in AUD and DD in most considered scenarios, respectively, while their CE performance is comparable to that of baseline algorithms.

\subsubsection{Performance Analysis for Different UE Activity Probabilities}
In Fig.~\ref{diff_alpha}, we present a comparative analysis of JACD performance across various UE activity probability scenarios, where the zero value of the UDER for Baseline 2, running 200 iterations under considered activity probabilities, is not shown on the logarithmic axis in Fig.~\ref{diff_alpha_UDER}. 
Firstly, Figs.~\ref{diff_alpha_UDER}-\ref{diff_alpha_ASER} illustrate a significant inverse relationship between UE activity probability and the JACD performance of different algorithms. 
Typically, with $200$ iterations, the proposed box-constrained FBS algorithm exhibits the best CE performance and achieves superior performance in AUD and DD. 
Conversely, with a reduced iteration count of $10$, the DU-ABC and DU-POEM algorithms generally outperform other benchmarks in AUD and DD despite their subpar CE performance, which is due to the exclusion of a metric assessing the channel estimation accuracy from the loss~function.

Analogous to the observations in Fig.~\ref{diff_P}, the DD of both DU-ABC and DU-POEM algorithms marginally decline at $\alpha=0.15$ relative to $\alpha=0.2$. 
This is because the loss function of our DU-based algorithms considers both AUD and DD performance, which might lead to the prioritization of one aspect over the other for a lower overall loss function value. 
Figs. \ref{diff_alpha_UDER} and \ref{diff_alpha_ASER} demonstrate the DU-based algorithms prioritize DD performance at higher probabilities ($\alpha=0.25, 0.3$) and AUD accuracy at lower probabilities ($\alpha=0.1, 0.15$).

In Fig.~\ref{ROC015}, we present the ROC comparison of various algorithms under $ \alpha = 0.15 $. 
Baselines 1 and 3 undergo $200$ iterations, while all other algorithms complete $10$ iterations each. 
The AMP-based Baseline 2 and Bi-GAMP-based Baseline 4 demonstrate the worst AUD performance. 
Moreover, our proposed DU-ABC and DU-POEM algorithms exhibit superior AUD performance, which attains the highest TPR for a given~FPR.

To sum up, when running $200$ iterations, the proposed box-constrained FBS algorithm typically achieves the best performance in AUD, CE, and DD. 
Conversely, the proposed DU-based algorithms generally surpass other baseline methods in AUD and DD performance with only $K=10$ iterations.

\section{Conclusions}\label{Section_VII}
We have proposed a novel framework of joint active user detection, channel estimation, and data detection (JACD) for massive grant-free transmission in cell-free wireless communication systems.
From this framework, we have developed several computationally efficient JACD algorithms, denoted as box-constrained FBS and PME-based JACD algorithms, accompanied by their deep-unfolded versions, DU-ABC and DU-POEM. 
When running $200$ algorithm iterations, the box-constrained FBS algorithm often exhibits superior JACD performance. 
When running only $10$ iterations, the proposed DU-ABC and DU-POEM algorithms usually significantly outperform all considered baseline methods regarding active user and data detection performance. 
The findings of this paper are expected to establish a solid foundation for the development of algorithms for massive machine-type communication.

\begin{appendices}
	\section{Proof of Proposition 1}\label{proof of Proposition 1}
As delineated in~\cite{sun2023joint_arxiv}, we recast the complex-valued optimization problem \eqref{xd_prox} into a real-valued problem:
\begin{equation}
	\begin{aligned}
		& \mathbf{r}_{n}^{(k+1)}\! = \!\!\!\!\mathop{\arg\min}_{\scriptstyle{\mathbf{r}_{n}^{(k+1)}\in\mathbb{R}^{2R_{\text{D}} }}}\frac{1}{2} \left\|\mathbf{r}_{n}^{(k+1)}\!-\!\hat{\mathbf{r}}_{n}^{(k)}\right\|_F^2+ \tau^{(k)}\mu_x\left\|\mathbf{r}_{n}^{(k+1)}\right\|_F\\
		& \text{s.t. } -B \le \mathbf{r}_{n}^{(k+1)}\!\left(d\right)\!\le B,\;\forall d\in\left\{1,\ldots,2R_{\text{D}}\right\},
	\end{aligned}
\label{r_n}
\end{equation}
where $\mathbf{r}_n^{(k+1)} \triangleq [\text{Re}\{\mathbf{x}_{\text{D},n}^{(k+1)}\}^T,\text{Im}\{\mathbf{x}_{\text{D},n}^{(k+1)}\}^T]^T\in\mathbb{R}^{2R_{\text{D}} }$ and $\hat{\mathbf{r}}_n^{(k)} \triangleq [\text{Re}\{\hat{\mathbf{x}}_{\text{D},n}^{(k)}\}^T,\text{Im}\{\hat{\mathbf{x}}_{\text{D},n}^{(k)}\}^T]^T\in\mathbb{R}^{2R_{\text{D}} }$. 
With the Lagrangian function $L(\mathbf{r}_{n}^{(k+1)},\mathbf{p},\mathbf{q})=\frac{1}{2} \|\mathbf{r}_{n}^{(k+1)}-\hat{\mathbf{r}}_{n}^{(k)}\|_F^2 +\tau^{(k)}\mu_x\|\mathbf{r}_{n}^{(k+1)}\|_F +  \sum_{d}\mathbf{p}(d)(\mathbf{r}_{n}^{(k+1)}(d)-B)- \sum_{d}\mathbf{q}(d)(\mathbf{r}_{n}^{(k+1)}(d)+B)$, the KKT conditions of optimization problem \eqref{r_n} are as follows~\cite{sun2023joint_arxiv}:
\begin{subequations}
	\begin{align}
		&\mathbf{r}_{n}^{(k+1)} + \frac{\tau^{(k)} \mu_x}{\left\|\mathbf{r}_{n}^{(k+1)}\right\|_F}\mathbf{r}_{n}^{(k+1)}-\hat{\mathbf{r}}_n^{(k)}+\mathbf{p}-\mathbf{q}=0,\label{40a}\\
		&\mathbf{r}_{n}^{(k+1)}(d)-B \le 0,\;-\mathbf{r}_{n}^{(k+1)}(d)-B \le 0,\;\forall d,\\
		&\mathbf{p}(d)\ge0,\;\mathbf{q}(d)\ge0,\;\forall d,\\
		&\mathbf{p}(d)(\mathbf{r}_{n}^{(k+1)}(d)-B)=0,\;\forall d,\\
		&\mathbf{q}(d)(\mathbf{r}_{n}^{(k+1)}(d)+B)=0,\;\forall d,
	\end{align}
\end{subequations}
where $\mathbf{r}_{n}^{(k+1)}\neq \mathbf{0}$.
The solution of equation \eqref{40a} is given by 
\begin{equation}
    \mathbf{r}_{n}^{(k+1)}\!=\!\frac{\max\!\left\{\!\left\|\hat{\mathbf{r}}_n^{(k)}\!-\!\mathbf{p}\!+\!\mathbf{q}\right\|_F\!\!-\!\tau^{(k)}\mu_x,0\right\}}{\left\|\hat{\mathbf{r}}_n^{(k)}\!-\!\mathbf{p}\!+\!\mathbf{q}\right\|_F}\left(\hat{\mathbf{r}}_n^{(k)}\!-\!\mathbf{p}\!+\!\mathbf{q}\right),
    \label{r_n_wo_0}
\end{equation}
with $\left\|\hat{\mathbf{r}}_n^{(k)}-\mathbf{p}+\mathbf{q}\right\|_F> \tau^{(k)}\mu_x$. 
Here, the expression \eqref{r_n_wo_0} implies that the influence of vectors $\mathbf{p}$ and $\mathbf{q}$ serves to diminish the values of some entries in $\hat{\mathbf{r}}_n^{(k)}$, thereby confining the elements of $\mathbf{r}_{n}^{(k+1)}$ to the interval $[-B,B]$ and, consequently, guaranteeing that $\mathbf{r}_{n}^{(k+1)}$ meets the KKT conditions.

To identify the set of elements in $\hat{\mathbf{r}}_n^{(k)}$ that potentially fall outside the interval $[-B,B]$, an initial resolution $\textstyle\mathbf{r}_{n,\text{tmp}}^{(k+1)}=\textsf{Shrinkage}\big(\hat{\mathbf{r}}_n^{(k)},\tau^{(k)}\mu_x\big)$ of equation \eqref{r_n} without constraints is conducted, i.e., $\mathbf{p}=\mathbf{q}=\mathbf{0}$, thereby we can obtain the sets $\mathcal{S}_p\triangleq \{d:\;\mathbf{r}_{n,\text{tmp}}^{(k+1)}(d)>B\}$ and $\mathcal{S}_q\triangleq \{d:\;\mathbf{r}_{n,\text{tmp}}^{(k+1)}(d)<-B\}$, which indicates the index of non-zero elements in $\mathbf{p}$ and $\mathbf{q}$, respectively. 
There are three cases as follows:
\begin{itemize}
	\item[1)] If $d\in\mathcal{S}_p$, then we should set $\mathbf{p}(d)>0$ and $\mathbf{q}(d)=0$ to reduce the proximal coefficient $\frac{\max\{\|\hat{\mathbf{r}}_n^{(k)}-\mathbf{p}+\mathbf{q}\|_F\!-\tau^{(k)}\mu_x,0\}}{\|\hat{\mathbf{r}}_n^{(k)}-\mathbf{p}+\mathbf{q}\|_F}$  and ensure the corresponding value of optimal vector  $\mathbf{r}_n^{(k+1)}(d)=B$.
	\item[2)] If $d\in\mathcal{S}_q$, then we should set $\mathbf{p}(d)=0$ and  $\mathbf{q}(d)>0$ to reduce the proximal coefficient $\frac{\max\{\|\hat{\mathbf{r}}_n^{(k)}-\mathbf{p}+\mathbf{q}\|_F\!-\tau^{(k)}\mu_x,0\}}{\|\hat{\mathbf{r}}_n^{(k)}-\mathbf{p}+\mathbf{q}\|_F}$ 
 and ensure the corresponding value of optimal vector  $\mathbf{r}_n^{(k+1)}(d)=-B$.
	\item[3)] If $d\notin\mathcal{S}_p\cup\mathcal{S}_q$, then we should set $\mathbf{p}(d)=0$ and $\mathbf{q}(d)=0$ because non-negativity of $\mathbf{p}$ and $\mathbf{q}$ ensures that the proximal coefficient $\frac{\max\{\|\hat{\mathbf{r}}_n^{(k)}-\mathbf{p}+\mathbf{q}\|_F\!-\tau^{(k)}\mu_x,0\}}{\|\hat{\mathbf{r}}_n^{(k)}-\mathbf{p}+\mathbf{q}\|_F}$ does not exceed $\frac{\max\{\|\hat{\mathbf{r}}_n^{(k)}\|_F-\tau^{(k)}\mu_x,0\}}{\|\hat{\mathbf{r}}_n^{(k)}\|_F}$, causing $|\mathbf{r}_n^{(k+1)}(d)|\le |\mathbf{r}_{n,\text{tmp}}^{(k+1)}(d)|$. Consequently, $\mathbf{r}_n^{(k+1)}(d)$ would still satisfy the conditions.
\end{itemize}

Given the index sets $ \mathcal{S}_p $ and $ \mathcal{S}_q $ corresponding to non-zero entries in vectors $ \mathbf{p} $ and $ \mathbf{q} $, we precisely determine the values of these non-zero entries. 
We now introduce the notation $ \mathbf{m} = \hat{\mathbf{r}}_n^{(k)} - \mathbf{p} + \mathbf{q} $, and $ b = \frac{{\max\{\|\mathbf{m}\|_F - \tau^{(k)}\mu_x, 0\}}}{{\|\mathbf{m}\|_F}} $ for brevity. 
Since $\left\|\mathbf{m}\right\|_F>\tau^{(k)}\mu_x$, then $b=\frac{\left\|\mathbf{m}\right\|_F-\tau^{(k)}\mu_x}{\left\|\mathbf{m}\right\|_F}\in\!\left(0,1\right]$ and we have $\mathbf{r}_{n}^{(k+1)}=b\,\mathbf{m}$, i.e., $\mathbf{m}(d)=\hat{\mathbf{r}}_n^{(k)}(d)\,\mathbb{I}\{d\notin\mathcal{S}_p\cup\mathcal{S}_q\}+\frac{B}{b}\,\mathbb{I}\{d\in\mathcal{S}_p\}-\frac{B}{b}\,\mathbb{I}\{d\in\mathcal{S}_q\}$.
Accordingly, $\left\|\mathbf{m}\right\|_F$ can be rewritten as $\|\mathbf{m}\|_F=\textstyle\sqrt{\textstyle{C_2 B^2}/{b^2}+C_1}$, where $C_1 = \sum_{d\notin \mathcal{S}_p\cup\mathcal{S}_q}\hat{\mathbf{r}}_n^{(k)}(d)^2$ and $C_2 = |\mathcal{S}_p\cup\mathcal{S}_q|$.
This can be substituted into $b=\frac{\left\|\mathbf{m}\right\|_F-\tau^{(k)}\mu_x}{\left\|\mathbf{m}\right\|_F}$ to obtain a quartic equation with respect to $b$ as:
\[
    2C_1 b^4 -4C_1 b^3 + (2C_1+C_2+C_3)b^2 -2C_2 b + C_2 = 0,
\]
where $C_3 = -2(\tau^d\mu_x)^2$.
Among the four solutions, the desired one lies within the range $(0,1]$.

If the aforementioned quartic equation has no solution in the range $(0,1]$, it means the case $\mathbf{r}_{n}^{(k+1)}\ne \mathbf{0}$ has no solution for problem \eqref{r_n}, then we can only consider $b=0$ and $\mathbf{r}_n^{(k+1)}=\mathbf{0}$.

This completes the proof.

\section{Proof of Proposition 2}\label{proof of Proposition 2}
Given observation vector $\hat{\mathbf{x}}=\mathbf{x} +\mathbf{e}$ with $\mathbf{x}\sim U_{\mathcal{S}}\{\mathbf{x}\}$ and Gaussian estimation error $\mathbf{e}\sim\mathcal{CN}\left(\mathbf{0},N_{\text{e}}\mathbf{I}\right)$, we can express $\text{PME}(\hat{\mathbf{x}},\mathcal{S},U_{\mathcal{S}}\{\mathbf{x}\},N_{\text{e}})$ as
\begin{equation}
     \text{PME}(\hat{\mathbf{x}},\mathcal{S},U_{\mathcal{S}}\{\mathbf{x}\},N_{\text{e}})=\frac{\sum_{\mathbf{x}\in\mathcal{S}}\mathcal{CN}(\mathbf{x};\hat{\mathbf{x}},N_{\text{e}}\mathbf{I})\mathbf{x}}{\sum_{\mathbf{x}\in\mathcal{S}}\mathcal{CN}(\mathbf{x};\hat{\mathbf{x}},N_{\text{e}}\mathbf{I})}.
\end{equation}
Meanwhile, for $\bar{\mathcal{S}}=\{\mathcal{S},\mathbf{0}\}$, we can express the PME of $\mathbf{x}$ under 
$U_{\alpha,\bar{\mathcal{S}}}\{\mathbf{x}\}$ as
\begin{equation}
    \begin{aligned}
     & \text{PME}(\hat{\mathbf{x}},\bar{\mathcal{S}},U_{\alpha,\bar{\mathcal{S}}}\{\mathbf{x}\},N_{\text{e}})\\
     &=\frac{\sum_{\mathbf{x}\in\mathcal{S}}\frac{\alpha}{|\mathcal{S}|}\mathcal{CN}(\mathbf{x};\hat{\mathbf{x}},N_{\text{e}}\mathbf{I})\mathbf{x}}{\sum_{\mathbf{x}\in\mathcal{S}}\frac{\alpha}{|\mathcal{S}|}\mathcal{CN}(\mathbf{x};\hat{\mathbf{x}},N_{\text{e}}\mathbf{I}) + (1-\alpha)\mathcal{CN}(\mathbf{0};\hat{\mathbf{x}},N_{\text{e}}\mathbf{I})}\\
     &=\frac{\alpha}{\alpha + (1-\alpha)\frac{|\mathcal{S}|\mathcal{CN}(\mathbf{0};\hat{\mathbf{x}},N_{\text{e}}\mathbf{I})}{\sum_{\mathbf{x}\in\mathcal{S}}\mathcal{CN}(\mathbf{x};\hat{\mathbf{x}},N_{\text{e}}\mathbf{I})\mathbf{x}}}\frac{\sum_{\mathbf{x}\in\mathcal{S}}\mathcal{CN}(\mathbf{x};\hat{\mathbf{x}},N_{\text{e}}\mathbf{I})\mathbf{x}}{\sum_{\mathbf{x}\in\mathcal{S}}\mathcal{CN}(\mathbf{x};\hat{\mathbf{x}},N_{\text{e}}\mathbf{I})}\\
     &=C_{\text{PME}}(\hat{\mathbf{x}},\mathcal{S},\alpha,N_{\text{e}})\,\text{PME}(\hat{\mathbf{x}},\mathcal{S},U_{\mathcal{S}}\{\mathbf{x}\},N_{\text{e}}),
\end{aligned}
\end{equation}
where
\begin{equation}
\!\!\textstyle C_{\text{PME}}(\hat{\mathbf{x}},\mathcal{S},\alpha,N_{\text{e}})\!\triangleq\!\alpha\left(\alpha\!+\!\left(1\!-\!\alpha\right)\!\frac{{|\mathcal{S}|}\,\mathcal{CN}(\mathbf{0};\hat{\mathbf{x}},N_{\text{e}}\mathbf{I})}{\sum_{\mathbf{x}\in\mathcal{S}}\mathcal{CN}(\mathbf{x};\hat{\mathbf{x}},N_{\text{e}}\mathbf{I})}\right)^{-1}\!\!\!.
\end{equation}
This completes the proof.

\section{Proof of Proposition 3}\label{proof of Proposition 3}
If $\mathcal{S}=\mathcal{R}^S$, we can rewrite $U_{\mathcal{S}}\{\mathbf{x}\}$ as
\begin{equation}
    \textstyle U_{\mathcal{S}}\{\mathbf{x}\}=\frac{1}{|\mathcal{R}^S|}=\frac{1}{|\mathcal{R}|^S}, \forall \mathbf{x}\in\mathcal{R}^S,
\end{equation}
which indicates entries of $\mathbf{x}$ are independent with each other.
Based on this, given observation vector $\hat{\mathbf{x}}=\mathbf{x} +\mathbf{e}\in\mathbb{C}^S$ with $\mathbf{x}\sim U_{\mathcal{S}}\{\mathbf{x}\}$ and Gaussian estimation error $\mathbf{e}\sim\mathcal{CN}\left(\mathbf{0},N_{\text{e}}\mathbf{I}_S\right)$, the $s$th entry of $\text{PME}(\hat{\mathbf{x}},\mathcal{S},U_{\mathcal{S}}\{\mathbf{x}\},N_{\text{e}})$ can be given by
\begin{equation}
    \begin{aligned}
        \textstyle\mathbf{e}_s^T\text{PME}(\hat{\mathbf{x}},\mathcal{S},U_{\mathcal{S}}\{\mathbf{x}\},N_{\text{e}})
        &=\frac{\sum_{\mathbf{x}\in\mathcal{S}}\frac{1}{|\mathcal{S}|}\mathcal{CN}(\mathbf{x};\hat{\mathbf{x}},N_{\text{e}}\mathbf{I}_S)\mathbf{e}_s^T\mathbf{x}}{\sum_{\mathbf{x}\in\mathcal{S}}\frac{1}{|\mathcal{S}|}\mathcal{CN}(\mathbf{x};\hat{\mathbf{x}},N_{\text{e}}\mathbf{I}_S)}\\
    &\overset{(a)}{=}\frac{\sum_{x\in\mathcal{R}}\frac{1}{|\mathcal{R}|}\mathcal{CN}(x;\hat{\mathbf{x}}(s),N_{\text{e}})x}{\sum_{x\in\mathcal{R}}\frac{1}{|\mathcal{R}|}\mathcal{CN}(x;\hat{\mathbf{x}}(s),N_{\text{e}})}\\
    &=\text{PME}(\hat{\mathbf{x}}(r),\mathcal{R},U_{\mathcal{R}}\{x\},N_{\text{e}}),
    \end{aligned}
\end{equation}
where $\mathbf{x}(s)=\mathbf{e}_s^T\mathbf{x}$ is replaced by $x$ in equation $(a)$. This completes the proof.
\end{appendices}

\section*{Acknowledgements}
The authors thank Victoria Palhares and Haochuan Song for their help in cell-free channel modeling and Gian Marti for his suggestions on deriving the FBS algorithm. We acknowledge Sueda Taner and Oscar Casta\~neda for their advice on training deep-unfolded algorithms. We are grateful to Mengyuan Feng for her suggestions on improving the figures in this paper. 

\ifCLASSOPTIONcaptionsoff
  \newpage
\fi

\balance
\bibliographystyle{IEEEtran}
\bibliography{./bib/Refabrv,./bib/IEEEBib1}
\balance

\end{document}